\documentclass[%
 reprint,
preprintnumbers,
 amsmath,amssymb,
 aps,
prb,
]{revtex4-2}
\usepackage{datetime2}
   \usepackage{xcolor}
\usepackage{amsmath}
\usepackage{subfigure}
\usepackage{graphicx}
\usepackage{dcolumn}
\usepackage{bm}
\usepackage{hyperref}
\usepackage{epstopdf}

\begin{document}

\title{Anomalous Inverse Spin Hall Effect (AISHE) due to Unconventional Spin Currents in Ferromagnetic Films with Tailored Interfacial Magnetic Anisotropy}

\author{Soumik Aon, Harekrishna Bhunia, Abu Bakkar Miah, Dhananjaya Mahapatra, Partha Mitra}

\email{pmitra@iiserkol.ac.in}

\affiliation{Department of Physical Sciences, Indian Institute of Science Education and Research, Kolkata, Mohanpur 741246, India}

\author{Pratap Kumar Pal, Anjan Barman}
\affiliation{Department of Condensed Matter Physics and Material Sciences, S. N. Bose National Centre for Basic Sciences,
Block JD, Sec. III, Salt Lake, Kolkata 700106, India}

\begin{abstract}
A single layer ferromagnetic film magnetized in the plane of an ac current flow, exhibits a characteristic Hall voltage with harmonic and second harmonic components, which is attributed to the presence of spin currents with polarization non-collinear with the magnetization. A set of 30 nm thick permalloy (Py) films used in this study are deposited at an oblique angle with respect to the substrate plane which induces an in-plane easy axis in the magnetization of the initial nucleating layers of the films which is distinct from the overall bulk magnetic properties of the film. This unusual magnetic texture provides a platform for the direct detection of inverse spin Hall effect in  Hall bar shaped macroscopic devices at room temperatures which we denote as Anomalous Inverse Spin Hall Effect (AISHE). Control samples fabricated by normal deposition of permalloy with slow rotation of substrate shows significant reduction of the harmonic Hall signal that further substantiates the model. The analysis of the second harmonic Hall signal corroborates the presence of spin-orbit torque arising from the unconventional spin-currents in the single-layer ferromagnets.   

\end{abstract}

\maketitle

\section{\label{sec:level1}Introduction}
A rare manifestation of relativistic effects in transport phenomenon are the spin-orbit induced Hall effects, namely spin Hall effect (SHE), anomalous Hall effect (AHE) and the inverse spin Hall effect (ISHE),  which are observable in any conductor with significant spin-orbit coupling strength (e.g. heavy metals). SHE refers to the phenomenon of generation of transverse spin current in response to an applied electric field (charge current) \cite{hirsch, Zhang,Kato_SHE_2004,valenzuela2006direct,hoffmann2013spin,sinova_2015_SHE}. For non-ferromagnetic heavy metals (HM), the transverse spin current is `pure' with no net charge transport associated with it and hence not detectable by direct electrical means. For ferromagnetic metals (FM), the transverse spin current is associated with a net charge current which leads to characteristic transverse voltage proportional to the magnetization of the FM, which is referred to as AHE \cite{Hall_1880,nagaosa2010anomalous}. ISHE \cite{saitoh_2006_ISHE,sinova_2015_SHE} is the phenomenon where a spin current flowing through a conductor results in a transverse charge current and hence a voltage. The microscopic mechanisms responsible for all the spin-orbit induced Hall effects are broadly classified into two categories, namely intrinsic mechanism that originates from band structure effects \cite{guo2008intrinsic} and extrinsic mechanisms that are caused by impurity/ defects scattering \cite{lowitzer2011extrinsic}. These mechanisms exhibit a characteristic relation that the  velocity vector of the carriers (current direction), the spin polarization and the transverse spin deflection direction forms a right handed coordinate system. The phenomenon of electrically generated spin currents, analysed  in a more general context \cite{davidson2020perspectives}, using an elegant argument that the symmetry conditions obeyed by the causes of a phenomenon  must also be preserved in the effects, leads to prediction of additional non trivial and interesting situations for the case of FMs. In the case of HM, an applied electric field say along the $x$-direction preserves the two mirror plane symmetry ($\sigma_{xy}$, $\sigma_{xz}$) and two fold rotation symmetry ($C^{x}_{2}$). Hence in this case the responses that preserve the said symmetries are  pure transverse spin currents with polarization perpendicular to both applied field and spin current direction, denoted as $Q_{zy}$ and $Q_{yz}$ where the current flow and polarization directions are indicated by the first and second indices in the subscript. Similar approach applied for the case of FM reveals that the presence of magnetization breaks additional symmetry which in turn creates more possibilities of spin current polarization compared to that of HM. In particular, if the magnetization is perpendicular to the electric field, say along the $y-$ axis, the $C^{x}_{2}$ and $\sigma_{xy}$ symmetries are broken that allow for a net charge current along the $z-$axis and a spin current $Q_{zy}$, which is the AHE. The symmetry argument further leads to nontrivial responses for more general situations. For the case where magnetization is exactly collinear with the applied electric field (current), all mirror plane symmetries are broken but $C^{x}_{2}$ symmetry is restored which rules out any transverse charge current, but the possibility of an unconventional spin current $Q_{zz}$ arises, such that the spin current is polarized along the flow direction along $z$-axis. Considering a specific but realistic experimental situation involving  `Hall-bar' devices of FM samples with strong in-plane magnetic anisotropy (e.g. Permalloy) with the applied current taken to be along the $x$-axis and the magnetization is confined in the $xy$-plane, a phenomenological expression for our devices for transverse spin current along $z$ in response to a charge current density ${J}_{C}$, is given as.
\begin{multline}\label{eq:1}
\vec{Q}_z = [\theta_{\parallel} cos(\phi)\hat{m} + \theta_{\perp}sin(\phi)\hat{m}_\perp + \theta_{\perp}^{R}sin(\phi) \hat{z}]{J}_{C}
\end{multline}
where, $\hat{m}$ and $\hat{m}_\perp$ are unit vectors parallel and  perpendicular to $\hat{m}$. 
The first term is identified as the longitudinal SHE with spins polarization colliner with the the magnetization with conversion efficiency or spin Hall angle $\theta_\parallel$. The second term is the transverse SHE with spin polarization perpendicular to magnetization but in plane with current and  $\theta_{\perp}$ being the corresponding spin Hall angle. The third term is identified as the SHE with rotation with spin polarization normal to both magnetization and current, $\theta_{\perp}^{R}$ being the corresponding spin Hall angle. The third term represents to the unconventional spin current $Q_{zz}$ for the specific case under consideration.

Experimental evidence of the self-induced SOT in FM metals has been shown through Spin-torque ferromagnetic resonance (ST-FMR) \cite{seki2021spin,fu2022observation}. Recently Wang et. al. \cite{wang2019anomalous} observed anomalous spin-orbit torque (ASOT) using MOKE in a single-layer FM, suggesting the presence of spin current with transverse spin polarization relative to the magnetization. Theoretically, Ryan et al. \cite{greening2022influence} supported this result by considering a nonuniform magnetization in a correspondingly thicker FM layer with a thickness exceeding a critical length known as the dynamic exchange coupling length. In these studies, the emergence of self-induced SOT is attributed to the symmetry breaking at the interface/surface \cite{wang2019anomalous,seki2021spin} or in the crystal \cite{Tang_2020_sym,Zhu_2021_sym}. When an in-plane current is applied in a FM layer, the current-induced SOT reorients the magnetization which can be detected as in-plane 2w harmonic Hall voltage \cite{avci2014interplay,du2021origin,greening2022influence}. Apart from that, few groups have recently reported the \cite{miao2013inverse,tsukahara2014self,azevedo2014addition} self-induced inverse spin Hall effect (ISHE) in permalloy (Py) which is caused by SOC. 


In our report, we conduct a thorough investigation of the first (1w) and second (2w) harmonic Hall measurements in a set of Py thin films deposited at an oblique angle, with variations of the in-plane component of the incoming flux. This unconventional deposition method induces an interfacial magnetic anisotropy, which is manifest in both the transport measurement and magneto-optical Kerr effect (MOKE) observations. The 1st harmonic measurement reveals two distinct contributions: (i) the symmetric contribution is attributed to the conventional planar Hall effect (PHE), while the asymmetric contribution is referred to as the anomalous inverse spin Hall effect (AISHE). The AISHE arises due to the conversion of the spin current into a charge current from the bulk/surface layer into the surface/bulk layer. Furthermore, we establish a correlation between the peak position of the asymmetric contribution and the angle of the incoming flux direction. This study allows us to determine the transverse spin Hall-like coefficient ($\theta^{R}_{\perp}$) for Py in these films. Additionally, we employ the 2nd harmonic Hall measurement technique to gain insights into the fundamental spin-orbit torque (SOT) phenomena using this series of devices. We have developed a toy model for both harmonics which confirms that spin current with polarization transverse to the magnetization exists in a FM. We quantitatively measure and disintegrate the current-induced SOT torques and their corresponding fields. From these measurements, we evaluate SOT efficiencies ($\xi_{\theta}$ and $\xi_{\phi}$) which help us to extract the spin polarization of Py films.

\begin{figure*}
\centering
\includegraphics[width=1.0\textwidth]{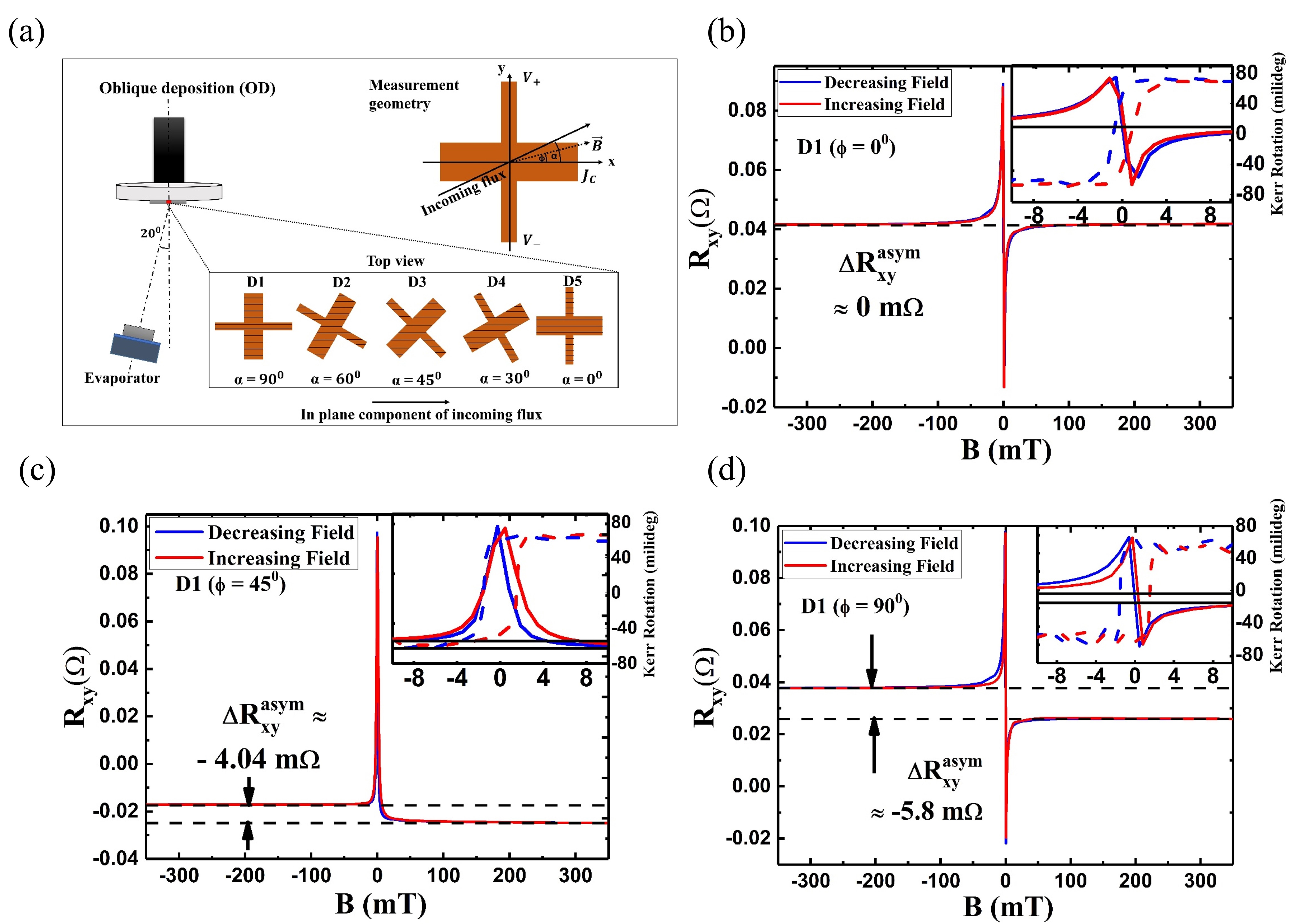}
\caption{(a) The schematic diagram portrays the fabrication method and device geometry. The incoming deposition flux forms an angle $\alpha$ with the x axis, varying within these devices as depicted. A current with a constant current density $J_{C}$ is applied along the x axis (main channel), while transverse voltage is measured along the y axis as an in-plane external magnetic field (B) rotates through a complete angle of $360^{0}$. Here, $\phi$ is defined as the angle between B and the x-axis. The transverse resistance denoted as $R_{xy}$, is computed using $R_{xy}$= ($V_{+}$ - $V_{-}$)/$J_{C}A$. (b), (c) and (d) shows $R_{xy}$ as a function of B, covering a range from -350 mT to +350 mT and vice versa. These measurements are conducted at $\phi$ = $0^{0}$, and $45^{0}$, and $90^{0}$, respectively, for device D1. Insets display both $R_{xy}$ and Kerr rotation vs B curves near zer field, ranging from -10 mT to +10 mT. The solid lines represent $R_{xy}$, while the dotted lines depict Kerr rotation. Blue and red data points signify the decrease and increase of B, respectively.} 
\label{Figure 1}
\end{figure*}

\subsection{\label{sec:level2}EXPERIMENTAL DETAILS}

We fabricated a series of Py devices where we could systematically vary the interfacial magnetization anisotropy by controlling the direction of incident Py flux on the device patterns while keeping the substrates fixed (no rotation during deposition). Several `Hall bar' patterns of dimensions 4 mm $\times$ 0.2 mm fabricated by photolithography were mounted on a planer substrate holder with different orientations of the current channel. The flux of Py was made to the incident at an oblique angle of $\sim 20^{0}$ with respect to normal to the substrate plane. We denote the angle made by the current channel of a particular Hall bar and the in-plane component of the Py flux direction as $\alpha$ as shown in Figure \ref{Figure 1}(a). We  report measurements on a set of five devices with $\alpha$ = $90^{0}$, $60^{0}$, $45^{0}$, $30^{0}$ and $0^{0}$ labeled as D1, D2, D3, D4, and D5 respectively. The thickness of Py grown by thermal evaporation (RADAK-I source from Luxel corporation) in all the devices $\sim$ 30 nm and the bulk characteristics like resistivity, anomalous Hall resistivity, planar Hall resistivity, X-ray spectrum,  morphology as measured in AFM did not show any pronounced variation. Kerr rotation measurements were performed in a dedicated static MOKE setup by scanning an in-plane magnetic field both parallel and perpendicular to the current channel that however revealed a clear variation in the in-plane magnetic anisotropy in the devices which as discussed in the following exhibits correlation with magneto transport measurement. The devices were mounted inside a four-pole electromagnet (Dexing, China), where the two perpendicular components of the magnetic field can be controlled independently such that the magnetic field vector of a certain magnitude can be swept in the plane of the substrate and the magnitude of the magnetic field can be varied in any particular direction in the plane. The devices were given an a.c. current excitation using a Keithley 6221A AC-DC current source and the corresponding first harmonic ($1\omega$) or second ($2\omega$) harmonic voltage responses were measured in the transverse direction using SR830 lock-in amplifiers. For Anomalous Hall Effect measurement, the samples were mounted differently so that the applied magnetic field varies normally to the substrate plane.\\

\begin{figure*}
\centering
\includegraphics[width=1.0\textwidth]{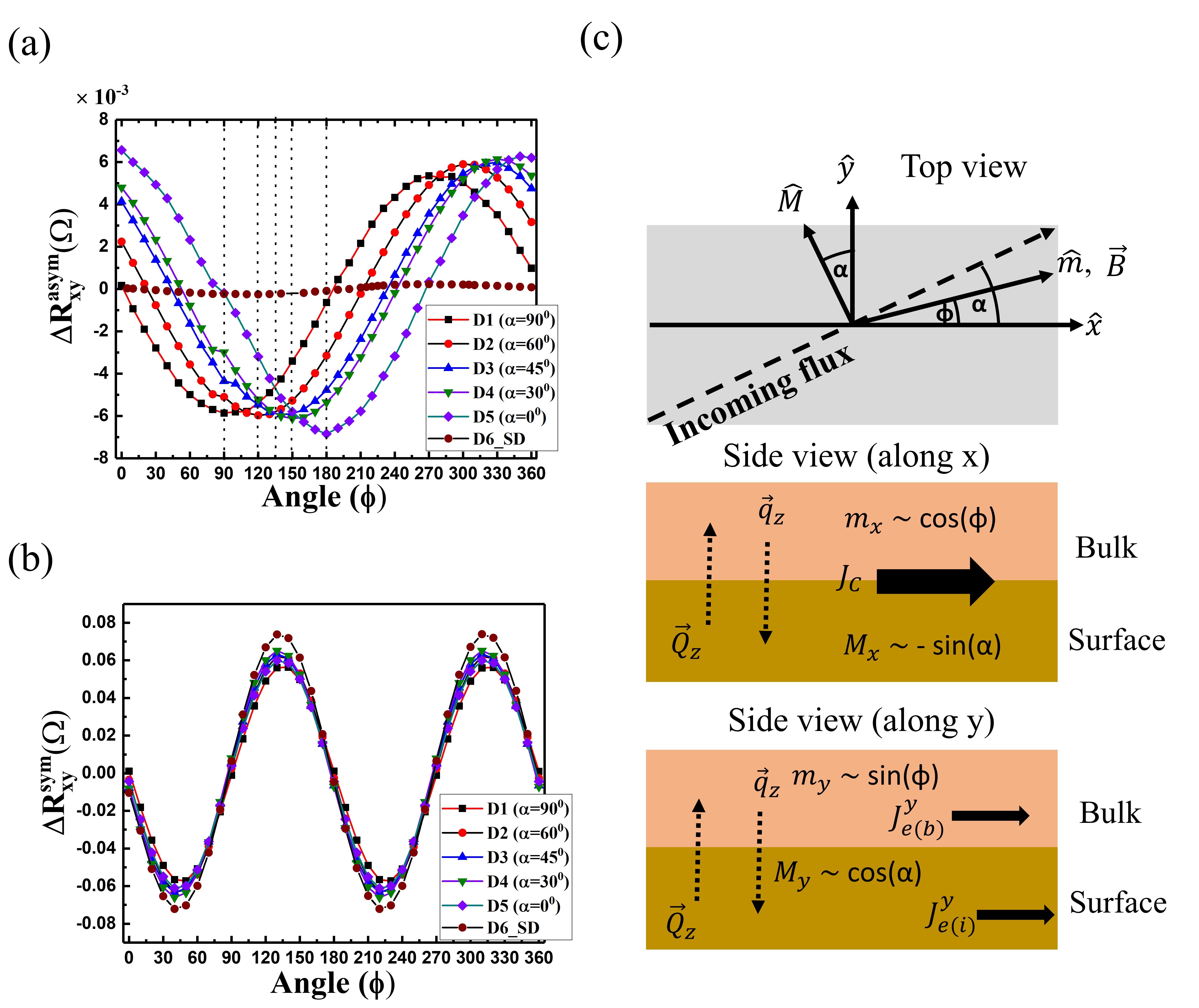}
\caption{(a) Angle dependence of asymmetric contribution $\Delta R^{asym}_{xy}$ is illustrated for all devices spanning from $\phi$=$0^{0}$ to $360^{0}$. A systematic shift in peak positions among the devices is observed, however, the peak amplitude remains nearly consistent across all devices, (b) The dominant influence of planar Hall effect is observed in the symmetric contribution $\Delta R^{sym}_{xy}$, fitted effectively with $sin(2\phi)$. Notably, $\Delta R^{sym}_{xy}$ displays no pronounced variation across different devices, (c) Depicts both the top and side view of a typical Py device. The bulk and surface magnetization are denoted by $\hat{m}$ and $\hat{M}$ respectively, while the incoming flux is indicated by the dashed grey arrow. The side view, along x and y axes, illustrates the flow spin current $q_{z}$ and $Q_{z}$ generated at bulk and surface layer, flowing along the -z and +z, respectively, and thereby producing a transverse charge current along y.}
\label{Figure 2}
\end{figure*}

\subsection{\label{sec:level3}RESULTS AND DISCUSSIONS}
\subsubsection{\label{sec:level1}{1st Harmonic Measurement}}
We begin with the description of $1\omega$ response for the devices. Figure \ref{Figure 1}(a)  shows the schematic of a device where the current and voltage leads of a typical Py Hall-bar are taken to be along x and y axes respectively and angle $\phi$ is defined as the angle between the x axis and applied in-plane magnetic field (B). For a fixed value of $\phi$  the in-plane magnetic field is scanned from 350 mT to -350 mT and to +350 mT in steps of 1 mT and the corresponding transverse resistances $R_{xy}=V^{1\omega}_{xy}/J_{C}A$ are recorded, where, $J_{C}$ is the current density defined, A is the cross-sectional area = $w$ $\times$ $t$ ($w$ and $t$ are width and thickness of the devices respectively). For all of our 30 nm Py devices, we find that the in-plane saturation fields as observed in MOKE signals and AMR measurements are found to be small  $B_{s}\sim 5$ mT, which is typical for Py. Therefore, for applied in-plane fields B $>$ 5 mT, the magnetization of the devices can be considered to be fully saturated along the direction of B. For ferromagnets with in-plane applied field and magnetization, transverse voltage is known to arise from the planar Hall effect \cite{Zhao_1997_PHE},\cite{Stavroyiannis_2003_PHE}, where $V_{xy}\propto m_{x}m_{y} \propto m_{s} sin(2\phi)$ such that the transverse voltage reaches maxima or minima at $\phi=45^{0}$, $135^{0}$ and zero at $\phi=0^{0}$, $90^{0}$, $180^{0}$ and $270^{0}$ for a fixed value of $m_s$. For a given $\phi$ the value of $R_{xy}$ is expected to remain constant for B $>$ $B_{s}$. Moreover, the planar Hall resistance is symmetric with respect to the reversal of magnetization direction for any angle $\phi$. Hence we expect that for B $>$ $B_{s}$, $R_{xy}$ should be the same for both positive and negative polarity of the applied field i.e. $R_{xy}({m})$ = $R_{xy}(-{m})$. However in all our devices depending on the scan angle $\phi$, we observe a characteristic $R_{xy}(B)$ curves that show pronounced anti-symmetric contribution such that $R_{xy}({m})$ $\neq$ $R_{xy}(-{m})$, indicating the presence of additional contribution to transverse voltage.

Figure \ref{Figure 1}(a)-(c) show the full scans of $R_{xy}(B)$ up to maximum applied fields of $\pm$ 350 mT for the sample D1 ($\alpha=90^{0}$) for  $\phi$ = $0^{0}$, $45^{0}$ and $90^{0}$ respectively. In the insets, the $R_{xy}(B)$ (solid lines) along with Kerr rotation curves (dashed lines) are plotted together for applied fields in the vicinity of switching fields between $\pm$ 10 mT, on the same scale of the y axis. The high field part of the $R_{xy}(B)$ curves are fitted to straight lines for $|B|\geq$ 100 mT which are extended over the entire scale. The difference of the intercepts of these lines at $B=0$ is denoted as $R_{xy}^{asym}$ a measure of the anti-symmetric component present in the measured $R_{xy}(B)$ data. This method of determining the anti-symmetric contribution ensures that any contribution from the ordinary Hall effect (OHE) or anomalous Hall effect (AHE) that may be present due to slight misalignment of the applied field from the substrate plane is nullified. It is worth emphasizing that Py is known to have strong in-plane anisotropy and it requires normal magnetic fields $\sim$ 0.8T to saturate the magnetization of the measured devices. Hence, a small out-of-plane component of the applied field may give rise to linear variation in anomalous Hall resistance which will be combined with the normal Hall effect. Figure \ref{Figure 1}(b), (c), and (d) show that for the device D1, $\Delta R_{xy}^{asym}$ $\approx$ 0, -4.04, and -5.8 $m\Omega$ for  $\phi$ = $0^{0}$, $45^{0}$, and $90^{0}$ respectively. $R_{xy}$(B) curves were recorded for $\phi$ varying from $0^{0}$ to $360^{0}$ in steps of $10^{0}$ and $\Delta R_{xy}^{asym}$ is calculated from each curve.  This process is repeated for other devices D2 to D5 and the result is plotted in Figure \ref{Figure 2} (a). We also calculate the symmetric component of the $R_{xy}(B)$ from the average of the two intercept values obtained from the linear fits of the high field parts, denoted as $\Delta R_{xy}^{sym}$. The results shown in Figure \ref{Figure 2}(b) indicate that the $\Delta R_{xy}^{sym}$ for all the devices under consideration (D1-D5) is essentially the manifestation of the planar Hall effect as expected, with zero value at four angular positions separated by $90^{0}$ and extreme values of same magnitude but opposite sign separated by angular position of $180^{0}$. We point out that the alignment of the current channel with the x component of the magnetic field is done visually under a microscope with an accuracy of $\sim5^{0}$. So we calibrate the angles $\phi$ by choosing the angles corresponding to maximum and minimum values of $\Delta R_{xy}^{sym}$ as $\phi$ = $45^{0}$ and $135^{0}$ respectively. Thus we can fit the data as $\Delta R_{xy}^{sym}=R_{PHE}sin(2\phi)$, and find the $R_{PHE}\sim 0.06 \Omega$, which is in good agreement with previously reported values \cite{elzwawy2021current}. In contrast, the variation of  $\Delta R_{xy}^{asym}$ with $\phi$ shows a systematic dependence on the deposition angle $\alpha$ of the devices. The data seems to exhibit a variation of the form $\Delta R_{xy}^{asym}=-R_{ISHE}sin(\phi-\phi_o)$, where the anti-symmetric component is maximum at $\phi$, and amplitude is denoted as $R_{ISHE}$. Our data [Figure \ref{Figure 2}(a)] shows an intriguing correlation that $\phi$ = $(180^{0}-\alpha)$ for all devices, within the experimental error in fixing the values of $\alpha$, which indicates that for a given device a magnetic anisotropy is developed such that the easy axis is along the deposition direction \cite{Zhou2018_anisotropy,Ali2021_anisotropy}. Thus when the magnetization is made to switch by scanning the field in that direction, we get maximum change in the transverse resistance. This additional anisotropy is possibly at the interfaces and the nature of bulk magnetization in the devices are same as reflected in the lack of any systematic variation in $\Delta R_{xy}^{sym}$ arising from PHE. Previous studies on electrical measurements on single-layer ferromagnets have reported the presence of an antisymmetric component in transverse resistance but ignore it as some sort of undesirable experimental artifact \cite{du2021origin}. However, we are able to show control over the variation of this antisymmetric component through our deposition technique. Furthermore, the nature of the $R_{xy}(B)$ curves near the switching fields is more intriguing.  The Kerr rotation curves are a measure of the surface magnetization as Py being a high permeability conductor the skin depth is small $\sim$ 12 nm \cite{wang2019anomalous}. As shown in Figure \ref{Figure 1}(b), (c), and (d), the Kerr rotation curves (dotted) show a typical hysteresis behavior indicative of strong in-plane anisotropy as expected in Py films.  However, for $\phi$ = $90^{0}$ the magnetization seems to switch abruptly at $B\sim\pm$ 2 mT, while at $\phi$ = $0^{0}$ the switching is more gradual and extends over the region  $\pm$ 4 mT. For $\phi$ = $45^{0}$, the nature of magnetization switching is in between the previous cases and the width of the hysteresis is larger. From the magnetization curves, as observed from Kerr rotation, we conclude that for $B\geq$ 5 mT the magnetization reaches the saturation value for any scan angle $\phi$. It is quite obvious that  $\phi$ = $90^{0}$ is an in-plane easy axis. But in that case, $\phi$ = $0^{0}$ should have exhibited the typical non-hysteresis `hard-axis' curve, which is not the case.  Analysis of the Kerr rotation data for device D5 (shown in Figure S2 in SM) reveals a complementary behavior to that of D1, i.e. $\phi$ = $0^{0}$ is the in-plane easy axis. Thus we observe a correlation between electrical and magnetic properties and the growth direction. The maximum of $\Delta R_{xy}^{asym}$ for a given device occurs when magnetization is switched along the easy axis of the device (determined by the angle $\phi$), which is determined by the incoming Py flux (angle $\alpha$).\\
Another interesting observation revealed in Figure \ref{Figure 1} is that the $R_{xy}$ is not directly following the variation in the magnetization for any given scan angle, which is also the trend in all devices. The blue lines are for B decreasing from 350 mT and the red line is for B increasing from -350 mT.  For $\phi$ = $0^{0}$ and $90^{0}$, we observe that $R_{xy}$ gradually increases, as the field is decreased from +350 mT and reaches a maximum at B $\rightarrow$ 0+ and then abruptly changes to a lower value as $\rightarrow$ 0- and then gradually increases as the field continues to decrease. A similar trend is observed when the field is gradually increased from -350 mT, with a small hysteresis. Thus there are two extrema in both increasing and decreasing scans, with the maxima appearing at B=0+ and the minima appearing at B=0- for both scans. The behavior for $\phi$ = $45^{0}$ is drastically different. For the decreasing field scan, the $R_{xy}$ gradually increases and reaches a minimum at B=0- and for the increasing field scan reaches a minimum at B=0+. Thus in both scans, there is only one extremum occurring at opposite polarity near B=0. This typical behavior of $R_{xy}$ near the switching fields at different scan angles, is universal for all the devices [see Figure S2 (device D5) in SM] and possibly originates from the bulk properties of the ferromagnet. As mentioned previously, the saturation value of $R_{xy}$(B) for large positive and negative fields depends on $\phi$ and hence $\Delta R_{xy}^{asym}$ is dependent on the scan angle and the device itself, indicating that it originates from interfacial properties of the ferromagnet.\\

Our data is an indication that there may be a gradient of magnetic properties along the thickness of the films which may be broadly classified as bulk and surface magnetization. Due to the finite skin depth of the incident laser light, the Kerr rotation arises primarily from surface magnetization and partly from bulk magnetization. Our current focus lies in uncovering the potential source of $\Delta R_{xy}^{asym}$ and it's connection to the anisotropy of the Py films. Recent studies \cite{davidson2020perspectives} indicate that for any arbitrary magnetization within a FM, spin current in the z-direction can be phonologically expressed as outlined in Eq. \ref{eq:1}. We have previously established that due to the oblique deposition, there could be a distinct difference in magnetization between bulk and surface layer. The unit vectors representing bulk and surface magnetization are symbolized by $\hat{m}$ and $\hat{M}$, respectively, as shown in Figure \ref{Figure 2}(c) (top view), where the dashed black arrow illustrates the incoming Py flux. The in-plane components of bulk and surface magnetization can be expressed as (see Note 3 in SM): $\hat{m} = [cos({\phi})\hat{x} + sin({\phi})\hat{y}]$ and $\hat{M} = [-sin({\alpha})\hat{x} + cos({\alpha})\hat{y}]$ as described in Figure \ref{Figure 2}(c) (side views). Here, we introduce a model aiming to explain origin of the asymmetric voltage resulting from the spin current in Py. This model distinguishes between the bulk and surface layers as distinct sources of spin currents. Considering the spin current $\Vec{q}_{z}$ generated in the bulk layer, it moves towards the -z direction and enters the surface layer. Similarly, the the spin current $\Vec{Q}_{z}$ originating from the surface layer flows towards z direction and enters the bulk layer. Given the material is single-layer, the spin transparency is expected be 100$\%$, indicating that no backward flow of spin current needs consideration. According to the Onsagar principle \cite{Onsagar_1931}, both these spin current along z axis generates charge currents along the y axis, which can be described as (see SM Note 3): \\

\begin{equation}\label{eq:4}
J_{e(asym)}^{y} = -(\theta_{\perp}^{R})^{2}J_{C}cos(\phi+\alpha)
\end{equation}
\noindent Py is known for its finite SOC, demonstrated by the observed self-induced inverse spin Hall effect (ISHE) in previous studies by various groups \cite{miao2013inverse,tsukahara2014self,azevedo2014addition}. In our investigation, the conversion of spin current to charge current between the bulk and surface layers is driven by the conventional ISHE mechanism. Despite the spin current in the FM layer differing from the typical spin current in non-magnetic (NM) layer, this phenomenon is termed the anomalous inverse spin Hall effect (AISHE). Due to this phenomenon, the transverse charge current is induced. To impede the flow of charge, an open circuit voltage arises, measurable, and describable using Eq. \ref{eq:4} as $\Delta V_{xy}^{asym}$ = $wE_{y}$ = $w\rho_{FM} J_{e(asym)}^{y}$ = -$w\rho_{FM} J_{C}(\theta^{R}_{\perp})^{2}cos(\phi+\alpha)$, where $w$ is the width of the Hall bar, $J_{C}$ is the applied current along the x-axis. The asymmetric Hall resistance $\Delta R_{xy}^{asym}$ is formulated as follows:\\

\begin{equation}\label{eq:5}
\Delta R_{xy}^{asym} = \Delta V_{xy}^{asym}/(J_{C}A) = -\frac{(\theta^{R}_{\perp})^{2} \rho_{FM} }{t} cos(\phi+\alpha)
\end{equation}
$\Delta R_{xy}^{asym}$ becomes maximum when $(\phi+\alpha)=180^{0}$, where the relation between $\phi$ and $\alpha$ are already established. From Eq. \ref{eq:5}, $\theta^{R}_{\perp}$ can be expressed as follows: $\theta^{R}_{\perp} = \sqrt{\frac{\Delta R_{xy}^{asym}(max)t}{\rho_{FM}}}$. For our devices, the $\theta^{R}_{\perp}$ is calculated considering the parameters: t = 30 nm and $\rho_{FM}$ = 71 $\mu \Omega$-cm. The magnitude of $\theta^{R}_{\perp}$ are estimated to be $\theta^{R}_{\perp}$ = 0.016 $\pm$ 0.001, 0.016 $\pm$ 0.0005, 0.016 $\pm$ 0.0006, 0.016 $\pm$ 0.0004, and 0.017 $\pm$ 0.001 for devices D1, D2, D3, D4, and D5 (see Table I in SM). The values are comparable with the reported experimental values for Py \cite{humphries2017observation,aljuaid_2018_spin_rotation}. The control device, fabricated using normal deposition technique with rotation, exhibits a significantly reduced $\Delta R_{xy}^{asym}$ value of approximately 0.24 $m\Omega$ [Figure \ref{Figure 2}(a)], negligibly smaller compared to those obliquely deposited devices. This conventional normal deposition technique might not induce the necessary interfacial magnetic texture, resulting in an undetectable voltage. This observation supports our proposed model.

\begin{figure*}
\centering
\includegraphics[width=1\textwidth]{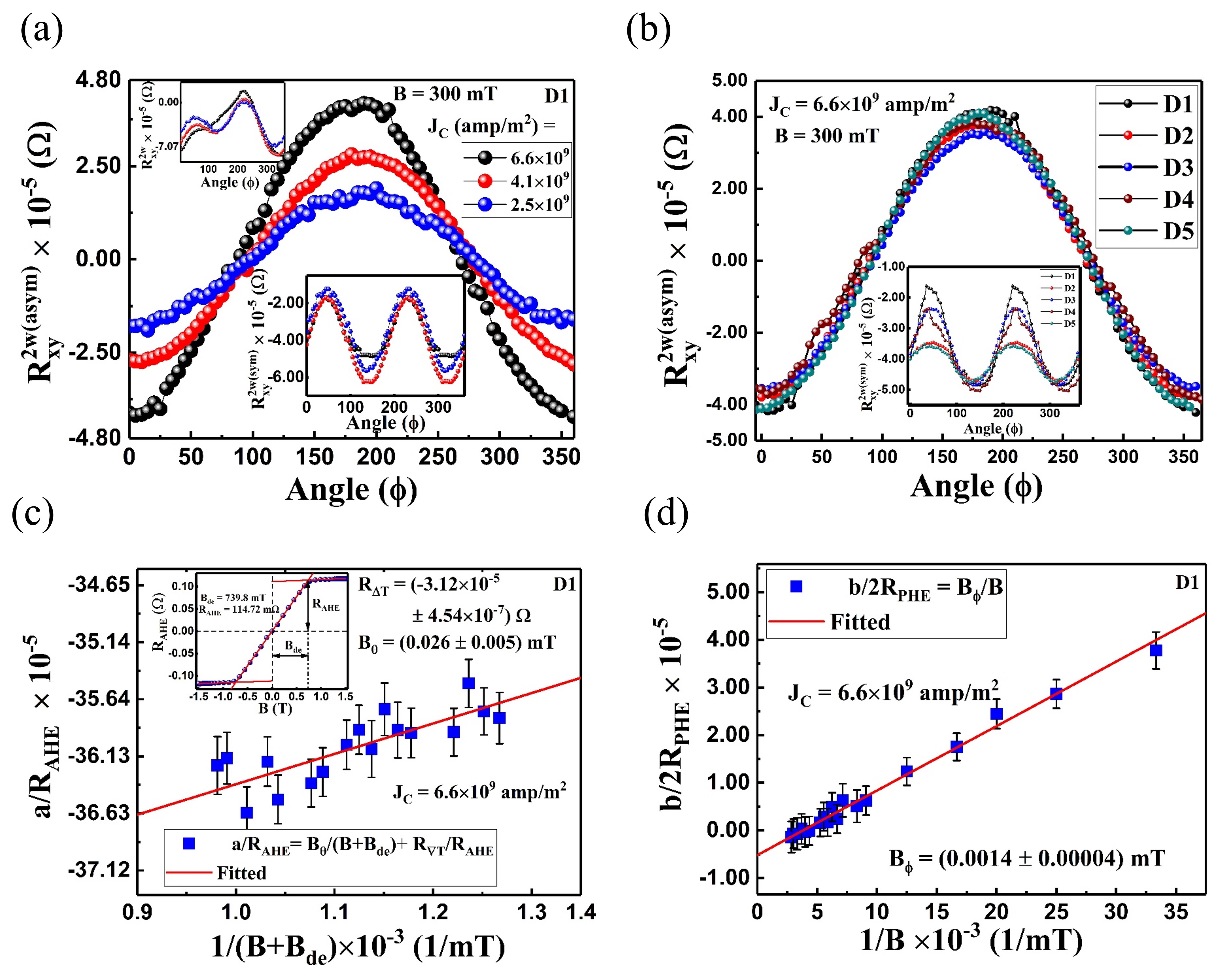}  
\caption{(a) The plot of $R^{2w(asym)}_{xy}$ against the angle $\phi$ is presented for $J_{C}$ = $2.5\times 10^{9}$, $4.1\times 10^{9}$, and $6.6\times 10^{9}$ $amp/m^{2}$, with B fixed at 300 mT for device D1. These curves are fitted using the equation $R^{2w(asym)}_{xy}$ = $a cos(\phi) + b cos(2\phi)cos(\phi) + c sin(\phi) +d cos(2\phi)sin(\phi)$ (Insets: $2w$ raw signal $R^{2w}_{xy}$ for $J_{C}$ = $2.5\times 10^{9}$, $4.1\times 10^{9}$, and $6.6\times 10^{9}$ $amp/m^{2}$ at B = 300 mT for device D1. The curves can be fitted using Eq. \ref{eq:6}. $R^{2w(sym)}_{xy}$ is depicted as a function of angle $\phi$ for $J_{C}$ = $2.5\times 10^{9}$, $4.1\times 10^{9}$, and $6.6\times 10^{9}$ $amp/m^{2}$ at B = 300 mT. The curves are fitted using $R^{2w(sym)}_{xy}$ = $e + fsin(2\phi)$, (b) The graph displays $R^{2w(asym)}_{xy}$ as a function of $\phi$ across all devices under $J_{C}$ = $6.6\times 10^{9}$ and B = 300 mT (inset: shows $R^{2w(sym)}_{xy}$ as a function of $\phi$ for all devices), (c) Shows the behaviour of $a/R_{AHE}$ vs $\frac{1}{(B+B_{de})}$. The curve is fitted with a straight line, the constant term yields the ANE term $R_{\nabla T}$, while the slope derives $B_{\theta}$ for device D1 (inset: $R_{AHE}$ is measured with an out-of-plane field up to 1.5 T. The resulting parameters are $R_{AHE}$ = 114.72 $m\Omega$ and $B_{de}$ = 739.8 mT. The blue circles represent the measured data points, and the red solid lines depict the fitted straight lines), (d) Illustrates $b/2R_{PHE}$ as a function of $\frac{1}{B}$. From the straight line fit, $B_{\phi}$ can be evaluated.}
\label{Figure 3}
\end{figure*}

\subsubsection{\label{sec:level2}SOT: 2nd Harmonic Measurement}

We explore self-induced SOT within single-layer of Py using current-induced 2nd harmonic in the same devices employed for the 1w measurements. Theoretical and experimental studies have confirmed that an in-plane current generates two distinct torques: damp-like (DL) torque and field-like torque, denoted by $\tau_{\theta}$ and $\tau_{\phi}$, respectively in a NM/FM bilayer \cite{garello2013symmetry,kim2013layer,avci2014interplay}. 
These torques can be described as follows: $\vec{\tau}_{\theta}$= $\tau_{\theta}$[$\hat{m}\times(\hat{m}\times\vec{Q}_{z})]$ and $\vec{\tau}_{\phi}$=  $\tau_{\phi}(\hat{m}\times\vec{Q}_{z})$, where, $\hat{m}$ signifies the magnetization unit vector within the FM layer and $Q_{z}$ represents the spin current generated by NM layer along z-axis. When the sample exhibits in-plane magnetization, DL and FL torques are accompanied by two distinct fields: $\vec{B}_{\theta}$=$B_{\theta}(\hat{m}\times\vec{Q}_{z})$ and $\vec{B}_{\phi}$=$B_{\phi}\vec{Q}_{z}$, respectively, where $\vec{B}_{\theta}$, $\vec{B}_{\phi}$ symbolizes the DL and FL fields. Recently, few groups have reported the experimental evidence of self-induced SOT within a single-layer FM, demonstrating the existence of both DL and FL torques \cite{seki2021spin,du2021origin,fu2022observation}. Nonetheless, the origin of these distinct torques within a single-layer FM remains a subject of ongoing debate. Our study delves into this by formulating the spin currents within the FM layer, and their respective torques, while also presenting the experimental validation.
Our approach is quite similar to the model employed for 1st harmonic transport experiment. The spin currents arise from both bulk and interface layers, exerting their respective torques on the surface and bulk layers (further detailed in Note 5 in SM). Therefore, the torques acting on the surface layer due to the bulk-layer spin current can be characterized as follows: $\vec{\tau}_{\theta}^{'}$ $\approx$ $\tau_{\theta}^{'}$ $[\hat{M}\times(\hat{M}\times\vec{q}_{z})]$ and $\vec{\tau}_{\phi}^{'}$ $\approx$ $\tau_{\phi}^{'}$ $(\hat{M}\times\vec{q}_{z})$, along with their corresponding fields $\vec{B}_{\theta}^{'}$ $\approx$ $B_{\theta}^{'}$ $(\hat{M}\times\vec{q}_{z})$ and $\vec{B}_{\phi}^{'}$ $\approx$ $B_{\phi}^{'}$ $\vec{q}_{z}$, where $\hat{M}$ represents the unit magnetization vector of surface, and $\vec{q}_{z}$ denotes the spin current generated at the bulk layer. Similarly, the torques acting on the bulk layer due to the surface-layer spin current can be described by: $\vec{\tau}_{\theta}^{''}$ $\approx$ $\tau_{\theta}^{''}$ $[\hat{m}\times(\hat{m}\times\vec{Q}_{z})]$ and $\vec{\tau}_{\phi}^{''}$ $\approx$ $\tau_{\phi}^{''}$ $(\hat{m}\times\vec{Q}_{z})$, along with their corresponding fields $\vec{B}_{\theta}^{''}$ $\approx$ $B_{\theta}^{''}$$(\hat{m}\times\vec{Q}_{z})$ and $\vec{B}_{\phi}^{''}$ $\approx$ $B_{\phi}^{''}$ $\vec{Q}_{z}$, where $\hat{m}$ represents the unit magnetization vector of bulk layer, and $\vec{Q_{z}}$ denotes the spin current generated at the surface layer. The resultant torques are denoted by $\tau_{\theta}$ and $\tau_{\phi}$ with their corresponding fields $B_{\theta}$ and $B_{\phi}$, respectively. When an ac current with a density of $J_{C}=J_{0}sin(wt)$ is applied in a FM, the magnetization can be deviated by the current-induced DL, and FL fields from its equilibrium position. The transverse Hall resistance undergoes oscillation at a frequency w, resulting in a 2nd harmonic component expressed by the equation (detailed in Note 5 in SM): 

\begin{multline}\label{eq:6}
R^{2w}_{xy}=a cos(\phi) +b cos(2\phi)cos(\phi)\\+c sin(\phi) + d cos(2\phi)sin(\phi) + e + f sin(2\phi)\\ 
\end{multline}

In our model, the coefficients c and d are significantly smaller compared to a and b. Consequently, for further spin-orbit field evaluation, we consider only the coefficients a and b. Thus, $R^{2w}_{xy}$ takes the following form:

\begin{multline}\label{eq:7}
R^{2w}_{xy} = (R_{AHE} \frac{B_{\theta}}{(B+B_{de})}+R_{\nabla T})cos(\phi)\\
                 +2R_{PHE} \frac{B_{\phi}}{B}cos(2\phi)sin(\phi)\\ + e +f sin(2\phi)
\end{multline}                 
In this equation, out of the plane ${B}_{\theta}$ contributes notably to the additional AHE in Hall measurements, while ${B}_{\phi}$ resides within the film plane, transverse to applied current, and modifies the PHE resistance. Here, $B$, $B_{de}$ represents applied, out-of-plane demagnetization field, respectively, and $R_{\nabla T}$ denotes the out-of-the-plane heating effect recognised as the anomalous Nernst effect (ANE). Due to significant conductivity variations between the substrate ($SiO_{2}$) and air, the in-plane current induces a perpendicular temperature gradient. This gradient dissipates heat through $SiO_{2}$, generating a perpendicular thermal gradient that manifests as the ANE \cite{avci2014interplay}. The in-plane Oersted field ($B_{Oe}$) generated by the FM layer maintains the symmetry with respect to the center of the FM layer and does not contribute in the Eq. \ref{eq:7}. Therefore, we can exclude the Oersted field $B_{Oe}$ from our calculation \cite{tshitoyan2015electrical}.\\

We conduct angular-dependence of $2w$ by applying a magnetic field B that rotates within the xy plane, spanning field ranging from 30 mT to 350 mT. The inset of Figure \ref{Figure 3}(a) shows  the raw $2w$ data $R^{2w}_{xy}$ at B = 300 mT and $J_{C}$ = $2.5\times 10^{9}$, $4.1\times 10^{9}$, and $6.6\times 10^{9}$ $amp/m^{2}$ for device D1, fitted using Eq. \ref{eq:6}. From the fitted curve, we extract coefficients a, b, c, d, e, and f (Eq. \ref{eq:6}), utilized in the subsequent analysis. These $2w$ data can be separated into asymmetric $R^{2w(asym)}_{xy}$ and symmetric $R^{2w(sym)}_{xy}$ contributions, previously discussed in the context of $1w$ measurements. [$a cos(\phi) +bcos(2\phi)cos(\phi)$] is designated as $R^{2w(asym)}_{xy}$, neglecting c and d as discussed earlier, while [$e + fsin(2\phi)$] constitutes $R^{2w(sym)}_{xy}$ which is considered as the symmetric contribution. In Figure \ref{Figure 3}(a), $R^{2w(asym)}_{xy}$  is displayed as a function of $\phi$ for current densities $J_{C}$ = $2.5\times 10^{9}$, $4.1\times 10^{9}$, and $6.6\times 10^{9}$ $amp/m^{2}$ at B = 300 mT. The inset of Figure \ref{Figure 3}(a) illustrates $R^{2w(sym)}_{xy}$, demonstrating the symmetric contribution. The increase in $J_{C}$ evidently amplifies the amplitude of $R^{2w(asym)}_{xy}$, which is consistent with the concept that the 2nd harmonic resistance arises from the SOT. However, $R^{2w(sym)}_{xy}$ could be perceived as a parasite contribution originating from the various sources, including potential misalignment between the device and B, discrepancies in the alignment of Hall bar voltage leads, and in-plane temperature gradient, a consequence of the device being warmer at it's center than its elongated edges \cite{avci2014interplay}. At $J_{C}$ = $2.5\times 10^{9}$, an accurate fitting of $sin(2\phi)$ is observed in $R^{2w(sym)}_{xy}$, however, for $J_{C}$ = $4.1\times 10^{9}$ and $J_{C}$ = $6.6\times 10^{9}$, slight deviations from $sin(2\phi)$ are observed in $R^{2w(sym)}_{xy}$. Considering that the other factors, such as voltage lead misalignment and the device orientation with respect to B, remain constant regardless of current variation, its plausible to claim that the higher current may induce pronounced in-plane heating, consequently causing the observed distortions. Figure \ref{Figure 3}(b) shows the $R^{2w(asym)}_{xy}$ for all devices with $J_{C}$ = $6.6\times 10^{9}$ at B = 300 mT. No significant change across the curves for different devices is observed and it is worth mentioning that the asymmetric data are these devices are dominated by the heating term $R_{\nabla T}$. The inset of Figure \ref{Figure 3}(b) showcases the plot of $R^{2w(sym)}_{xy}$, showing slight fluctuations among the different devices. However, no systematic pattern is observed in the variation among these curves, highlighting various potential misalignment previously discussed. The damp-like and filed-like terms can be quantitatively identified by magnetic field dependence as presented in Figure \ref{Figure 3}(c) and (d) for device D1. The coefficients $a = (R_{AHE} \frac{B_{\theta}}{(B+B_{de})}+R_{\nabla T})$ and $b = 2R_{PHE} \frac{B_{\phi}}{B}$, described in Eq. \ref{eq:7} are separated. Earlier, $R_{PHE}$ is estimated to be $\approx$ 0.06 $\Omega$. To determine $R_{AHE}$, an out-of-plane B ranging from -1.5 T to +1.5 T is swept, as shown in the inset of Figure \ref{Figure 3}(c). At higher fields, the ordinary Hall effect (OHE) starts to dominate,
necessitating elimination by linearly fitting the data points at the higher field range. The blue solid circles represent the measured data points, while the red solid lines depict the fitted straight lines. The evaluated parameters are $R_{AHE}$ = 114.72 $m\Omega$ and $B_{de}$ = 739.8 mT. In experiments and theoretical considerations of a FM/NM bilayer, $B_{\theta}$ emerges from the bulk spin Hall effect of NM layer. Moreover, the independence of $B_{\phi}$ on thickness leads to the inference that the interface serves as the origin for $B_{\phi}$ in a FM/NM bilayer. Additionally, the Rashba effect at the surface/interface stands as the another plausible contributor to $B_{\phi}$. In our experiment, our focus lies in the qualitative exploration of the spin-orbit effects within the Py single-layer films with various magnetic anisotropy at the interfaces. The objective is not centered on investigating the quantitative influence of different spin-orbit fields, which would require a study dependent on thickness variations \cite{du2021origin}. Here, $B_{\theta}$ can be evaluated by plotting $a/R_{AHE}$ as a function of $\frac{1}{B+B_{de}}$ for $J_{C}$ = $6.6\times10^{9} amp/m^{2}$. Through fitting a straight line to this curve, $B_{\theta}$ is determined from the slope, with the intercept represents $R_{\nabla T}/R_{AHE}$. The resulting $B_{\theta}$ is calculated as (0.026 $\pm$ 0.005) mT and $R_{\nabla T}$ is -(3.12$\times 10^{-5}$ $\pm$  4.54$\times 10^{-7}$) $\Omega$ for device D1. Similarly, using the $b/2R_{PHE}$ vs $1/B$ curve facilitates the computation of the FL field, yielding $B_{\phi}$ = (0.0014 $\pm$ 0.00004) mT for device D1. We repeat the experiments to determine the $B_{\theta}$ and $B_{\phi}$ fields for all devices. The resulting values of $B_{\theta}$ are (0.024 $\pm$ 0.001) mT, (0.010 $\pm$ 0.003) mT, (0.019 $\pm$ 0.002) mT, and (0.020 $\pm$ 0.005) mT, while $B_{\phi}$ measures (0.00085 $\pm$ 0.00004) mT, (0.0012 $\pm$ 0.00003) mT, (0.0015 $\pm$ 0.00005) mT, and (0.0024 $\pm$ 0.00005) mT for devices D2, D3, D4, and D5 respectively (see Table 2 in SM). The measured values of $B_{\theta}$ in our experiments significantly exceed those of Py single-layer caped with $Al_{2}O_{3}$ as reported in Seki et al. \cite{seki2021spin}. The difference may come from the notably greater thicknesses of our devices, allowing for increased spin current generation, resulting in a higher torque, as demonstrated in Du et al. \cite{du2021origin}. The efficiency of SOT torque can be characterized by \cite{Khvalkovskiy_2013_efficiency,pai2015dependence}

\begin{equation}\label{eq:8}
\xi_{\theta(\phi)} = \frac{2e}{\hbar} \frac{\mu_{0}t_{Py}B_{\theta(\phi)}M_{S}}{J_{C}}    
\end{equation}
where, e is the electron charge, $\hbar$ is Dirac constant, $M_{S}$ is the saturation magnetization of Py film and $\xi_{\theta(\phi)}$ is SOT efficiency of DL(FL) torques \cite{Pai_2014_efficiency}. $M_{S}$ is taken $\approx$ 85.8 mT from \cite{pal2022short} and for $J_{C}$ = $6.6\times10^{9} amp/m^{2}$, we estimate $\xi_{\theta}$ = (0.024 $\pm$ 0.003), (0.022 $\pm$ 0.0009), (0.009 $\pm$ 0.003), (0.018 $\pm$ 0.002), and (0.019 $\pm$ 0.005), and $\xi_{\phi}$ = (0.0014 $\pm$ 0.00004), (0.0008 $\pm$ 0.00004), (0.0011 $\pm$ 0.00003), (0.0014 $\pm$ 0.00005), and (0.0022 $\pm$ 0.00005) for devices D1, D2, D3, D4, and D5, respectively. The efficiency of SOT and the effective spin Hall angle of FM material is related through the subsequent equation:

\begin{equation}\label{eq:9}
\xi_{\theta(\phi)} = \theta^{eff}_{SHE} (1-sech(\frac{t_{Py}}{\lambda_{Py}}))
\end{equation}

where, $\theta^{eff}_{SHE}$ represents the effective spin Hall angle of Py, $t_{Py}$ and $\lambda_{Py}$ are thickness and spin diffusion length of Py. We deliberately select a substantial $t_{Py}$ (30 nm) to fulfill the condition ($t_{Py}$ $\ge$ $\lambda_{Py}$), resulting in $sech(\frac{t_{Py}}{\lambda_{Py}})$ $\approx$ zero. Therefore, in our experiment $\xi_{\theta}$ is nearly equal to $\theta^{eff}_{SHE}$. In our measurements on the series of Py devices, the observed effective spin Hall angle is comparable to the order of magnitude of the spin Hall angle like efficiency of the ASOT reported by Wang et al. \cite{wang2019anomalous}. However the magnitude we obtain in our experiments is notably lower than that observed in the ASOT experiment. In a FM $\theta^{eff}_{SHE}$ correlates to $\theta_{AHE}$ through the relationship $\frac{\theta^{eff}_{SHE}}{\theta_{AHE}}= \frac{1}{P_Py}$ \cite{tsukahara2014self}, where $\theta_{AHE}$ signifies the anomalous Hall angle and $P_{Py}$ is the spin polarization of Py. $\theta_{AHE}$ is defined by $\theta_{AHE}$ = $\frac{\rho_{AHE}}{\rho_{xx}}$, where $\rho_{AHE}$ and $\rho_{xx}$ denote anomalous Hall resistivity and longitudinal resistivity. The measured $\rho_{xx}$ is roughly 71 $\mu\Omega-cm$ for our devices. We determine that the average $\theta_{AHE}$ = 0.005 for Py films. By averaging the $\theta^{eff}_{SHE}$ values across the devices, and employing the $\theta^{eff}_{SHE}$ and $\theta_{AHE}$ relation, we obtain the spin polarization $P_{Py}$ $\approx$ 0.25, which is comparable to the values obtained utilizing the lateral spin-valve structure \cite{sagasta2017spin,omori2019relation}.

\subsection{\label{sec:level5}SUMMARY}
To summarize, we have fabricated a series of Py Hall bars (30 nm) using a fixed angle oblique deposition technique, giving us control over the in-plane incoming material flux of Py. This unusual deposition method induces surface magnetic anisotropy distinct from the bulk of the films, resulting in a detectable voltage generated due to unconventional spin current generated within the FM single-layers. We have developed a toy model based on the the generalised formula for generated spin currents and their conversion with the charge current within a FM material. Our proposed model has been validated through electrical measurements employing both 1st and 2nd harmonics. The 1st harmonic measurement, conducted under an in-plane applied magnetic field, captures the asymmetric transverse voltage $R^{2w(asym)}_{xy}$, displaying a sinusoidal relationship with angle $\phi$. Our proposed model suggests that due to the distinct magnetic texture in the surface layer compared to the bulk layer in different devices, the spin currents generated from both layers penetrate the respective layers, leading to the self-induced anomalous inverse Hall effect (AISHE) within the Py films. This experimental measurement aligns with the proposed model. Moreover, this observation, in conjunction with MOKE measurement, establishes a link between the angle $\phi$ at maximum $R^{2w(asym)}_{xy}$ and their in-plane incoming flux angle $\alpha$. The devices peak at the angle where the magnetization aligns with the soft axis. Additionally, the transverse spin Hall like coefficient is evaluated, confirming the existence of spin currents  transverse to the magnetization within a FM. The 2nd harmonic measurement effectively explain the conventional self-induced spin-orbit torque (SOT) by analysing the damp-like (DL) and field-like (FL) torques, along with their corresponding fields $B_{\theta}$ and $B_{\phi}$, respectively. Employing a similar formulation for calculating the torques originating from the spin current generated by both bulk and surface layers and their respective fields, exhibits strong agreement with the experimental results. The SOT efficiency $\xi_{\theta(\phi)}$ and the effective spin Hall angle $\theta_{SHE}^{eff}$ is evaluated for measured Py films. From the relationship between $\theta_{AHE}$ and $\theta^{eff}_{SHE}$, we evaluate the spin polarization $P_{Py}$ of Py, comparable to values observed in the non-local spin valve (NLSV) devices. Hence, these simple yet efficient devices significantly contribute to apprehending the spin-related phenomena in FM materials.\\

\subsection{\label{sec:level6}ACKNOWLEDGEMENT}
We acknowledge the Ministry of Education (MoE), Government of India and IISER Kolkata for providing the fellowship and necessary funding.

\clearpage  
\onecolumngrid  
\noindent {\textbf{\LARGE Supplementary Material for \\`Anomalous Inverse Spin Hall Effect (AISHE) due to Unconventional Spin Currents in Ferromagnetic Films with Tailored Interfacial Magnetic Anisotropy'}\\

\setcounter{figure}{0}
\renewcommand{\thefigure}{S\arabic{figure}}

\clearpage

\noindent \textbf{Note 1. Characterization of Py films}:\\

\noindent \textbf{Note 2. MOKE and transport measurement ($1w$) of device D5}:\\

\noindent \textbf{Note 3. Theoretical formulation describing the origin of $\Delta R_{xy}^{asym}$ (1st harmonic transport measurement)}:\\

\noindent \textbf{Note 4. 2w field scans for devices D1 and D5}:\\

\noindent \textbf{Note 5. Calculation for 2w transport measurement}:\\

\clearpage

\noindent \textbf{Note 1. Characterization of Py films}:\\
\begin{figure}[h!]
\centering
\includegraphics[width=1\textwidth]{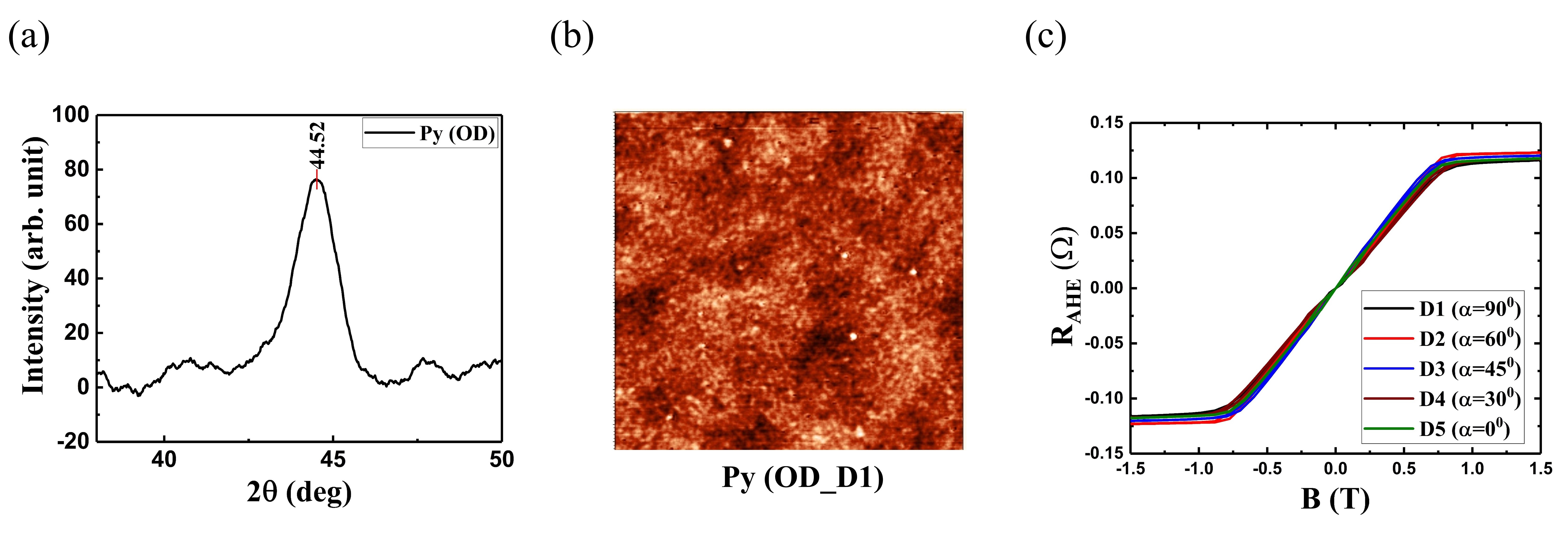}
\caption{(a) XRD image of obliquely deposited Py thin film of thickness 30 nm. The observed peak at $2\theta$ = $44.52^{0}$ (OD) confirms the presence of fcc crystal structure, (b) AFM image captures the surface of device D1, covering a scanning area of 2 $\mu m$ $\times$ 2 $\mu m$, (c) AHE data are collected for all devices, indicating nearly identical magnitudes.}
\label{SM_Figure 1}
\end{figure}

\noindent \textbf{Note 2. MOKE and transport measurement ($1w$) of device D5}:\\

\begin{figure}[h!]
\centering
\includegraphics[width=1.0\textwidth]{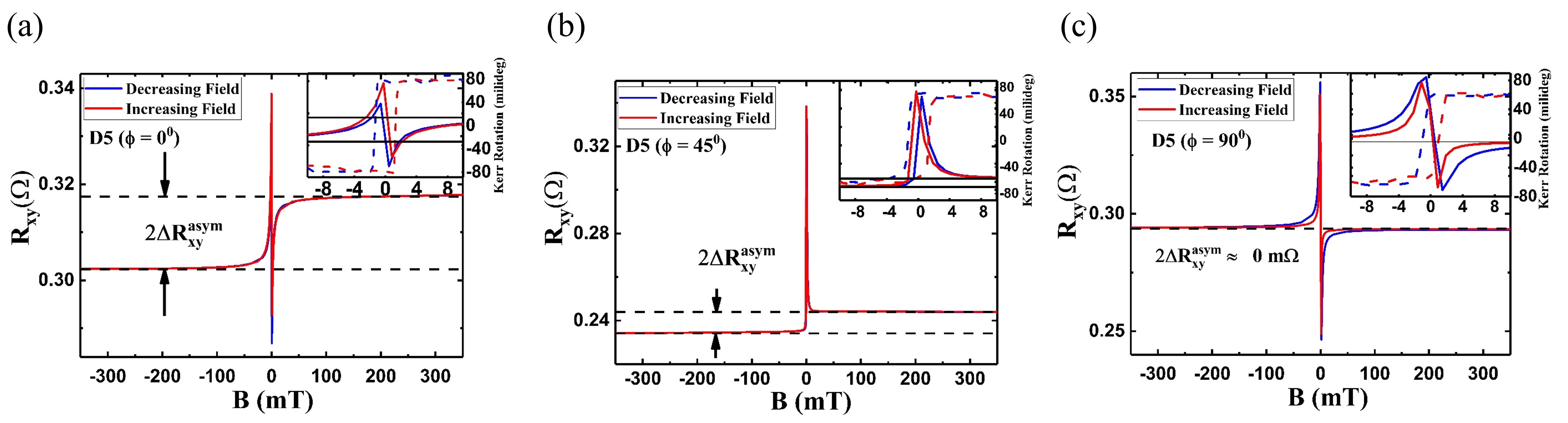}
\caption{(a), (b) and (c) Shows $R_{xy}$ as a function of B, covering a range from -350 mT to +350 mT and vice versa. These measurements are conducted at $\phi$ = $0^{0}$, and $45^{0}$, and $90^{0}$, respectively, for device D5. Insets display both $R_{xy}$ and Kerr rotation as a function of B near zero field, ranging from -10 mT to +10 mT. The solid lines represent $R_{xy}$, while the dotted lines depict Kerr rotation. Blue and red data points indicate the decrease and increase of B, respectively.}
\label{SM_Figure 2}
\end{figure}
\clearpage

\noindent \textbf{Note 3. Theoretical formulation describing the origin of $\Delta R_{xy}^{asym}$ (1st harmonic transport measurement)}:\\

\begin{figure}[h!]
\centering
\includegraphics[width=1.0\textwidth]{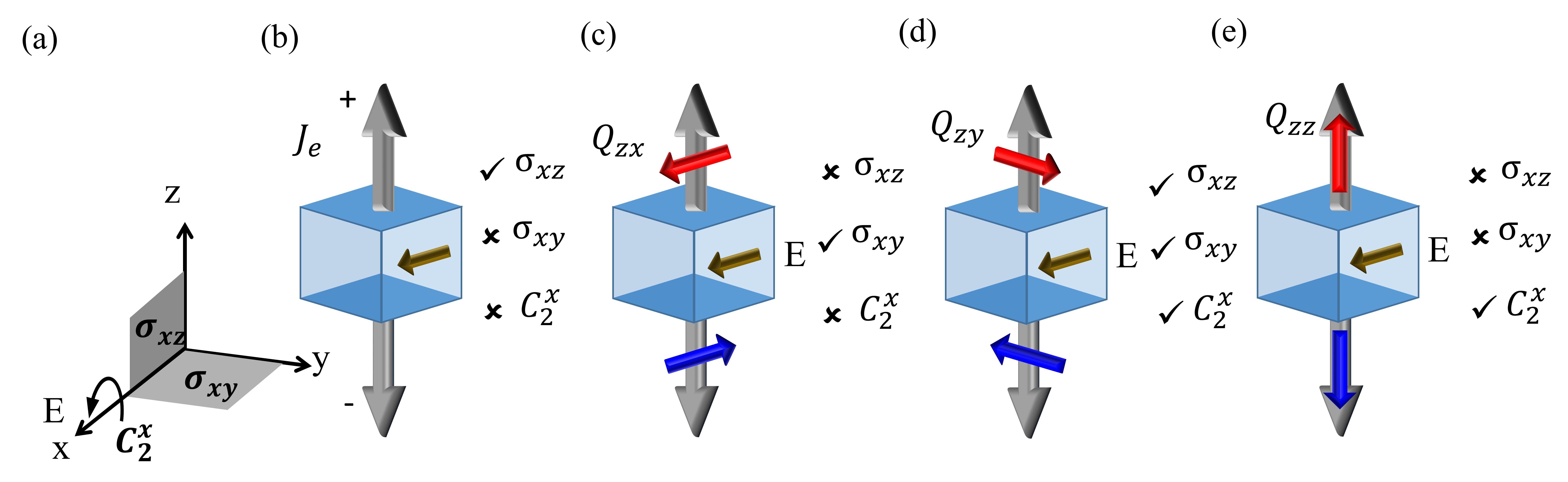}
\caption{Illustration of the permissible spin currents that are generated electrically in a NM material. (a) Depicts the standard coordinate system, with electric field (E) applied along x-axis, (b)-(e) Describe both the effect of E (on the left) and the protected or violated symmetries (on the right). The dark gold arrow represents E, while the red and blue arrows symbolize two spin orientations. In (b), grey arrow represents the direction of $J_{e}$, and in (c)-(e) it symbolizes the flow of spin currents. Notably, applied E does not generate any charge current along z-direction. (c)-(e) Show spin currents along z-axis with spin polarization in the x,y, and z-axes. The sole possible spin current flowing along the z-axis with spin polarization in the y direction is represented as $Q_{zy}$.}
\label{SHE_symmetry}
\end{figure}
We can explain the SHE by considering the symmetry in crystal. According to Curie's principle \cite{Curie_1970}, the cause and effect of an event should preserve the same symmetry. Figure \ref{SHE_symmetry} illustrates the allowed spin current in SHE in a NM material, using the symmetry argument. Here, electric field (E) is the cause and consequent generation of spin current represents the effect. In a cubic crystal, three types of symmetries are present: (i) bulk inversion symmetry, (ii) rotational symmetry around the x, y, and z-axes ($C^{x}_{2}$, $C^{y}_{2}$, $C^{z}_{2}$ respectively), (iii) mirror symmetry across xy, yz, and xz plane ($\sigma_{xy}$, $\sigma_{yz}$, and $\sigma_{xz}$, respectively). When E is applied along x axis, inversion symmetry is broken and rotational symmetry along y, z ($C^{y}_{2}$, $C^{z}_{2}$) are eliminated. Also the mirror symmetry about the xz plane ($\sigma_{xz}$) is removed. However, $C^{x}_{2}$, $\sigma_{yz}$, and $\sigma_{xy}$ remain preserved. Now, our objective is to understand the possible charge and spin currents that maintain these symmetries. The generated charge current $J_{e}$ along z and spin currents $Q_{zx}$, $Q_{zz}$ do not protect these symmetries, so they are not allowed. Only $Q_{zy}$ complies with the necessary symmetries and remains permissible.

\begin{figure}
\centering
\includegraphics[width=1.0\textwidth]{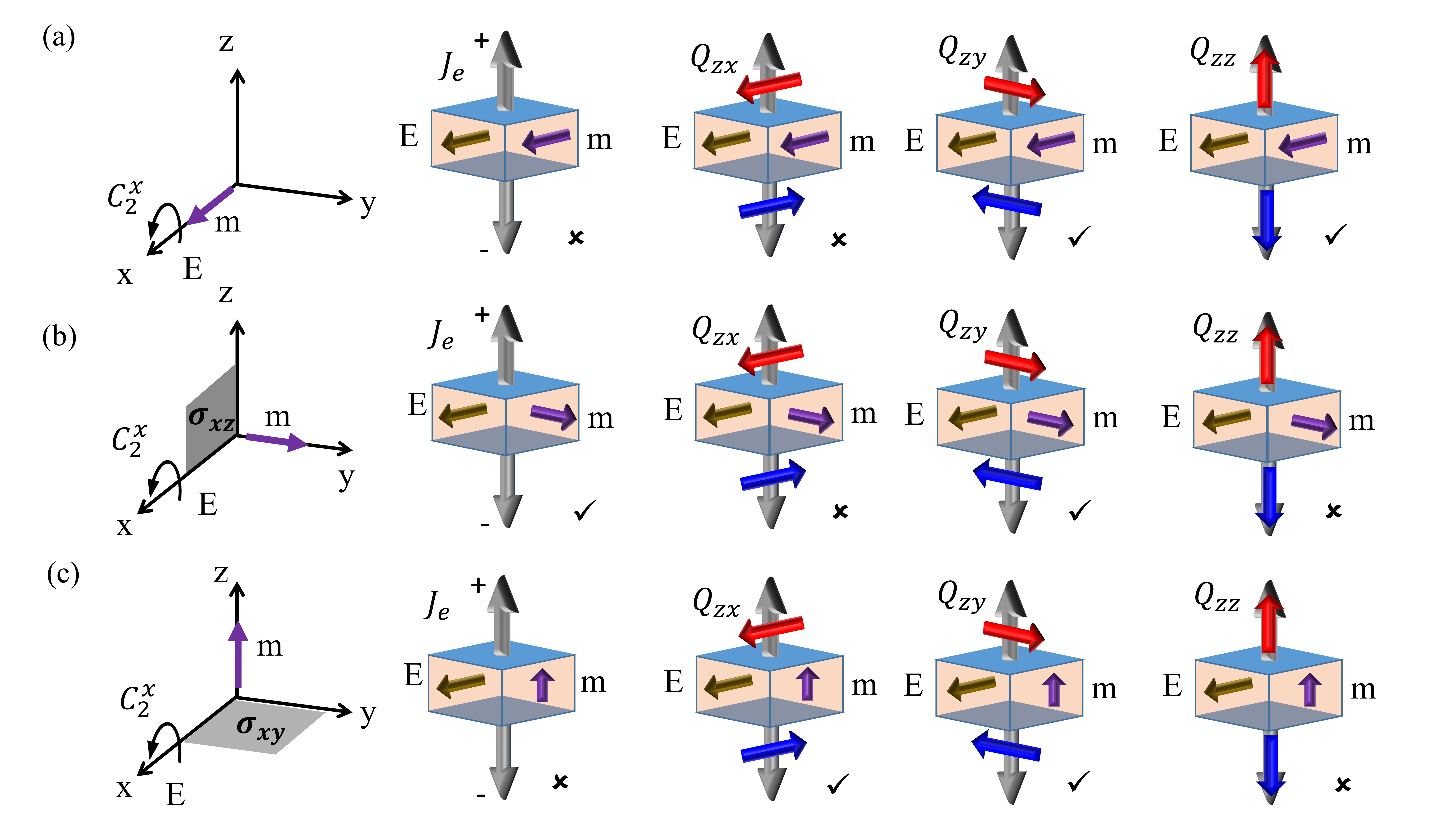}
\caption{Illustration of electrically generated spin currents within a FM material, where magnetization aligns with (a), (b), (c) x, y, and z-axes, respectively. The electric field is applied along x-axis. The sole allowed charge current is $J_{C}$, with magnetization oriented along y-direction. The permissible spin currents are $Q_{zx}$, $Q_{zy}$, and $Q_{zz}$.}
\label{FM_Symmetry}
\end{figure}
The same symmetry argument gives rise to the anomalous Hall effect in ferromagnetic material. In a ferromagnet, the magnetization breaks symmetries which lifts off the further constraint of allowed spin current along with electric field. Mirror symmetries about the plane that contain the magnetization and rotational symmetries about the axes perpendicular to the magnetization can be broken in the presence of magnetization. Figure \ref{FM_Symmetry} shows the allowed spin currents in the presence of an electric field and magnetization. When magnetization is directed along y-direction, depicted in figure \ref{FM_Symmetry}(b), charge current $J_{e}$ is generated that refers to the conventional AHE. Other permissible spin currents include $Q_{zy}$, $Q_{zz}$ (magnetization along x), $Q_{zy}$ (magnetization along y), $Q_{zx}$, $Q_{zy}$ (magnetization along z). Hence, the spin currents transverse to the magnetization exist in FM materials.\\

Based on the symmetry argument as described by Dadivson et al. \cite{davidson2020perspectives}, we can formulate a generalized expression for the spin current flowing in the z-direction within a ferromagnet (FM) as follows: 
\begin{equation}\label{eq:1}
\vec{Q}_z = \sigma_{\parallel}[\hat{m}\cdot(\hat{z}\times \vec{E})]\hat{m} + \sigma_{\perp}\hat{m}\times [(\hat{z}\times \vec{E})\times \hat{m}] + \sigma_{\perp}^{R} [\hat{m}\times (\hat{z}\times \vec{E})]
\end{equation}
where, $\sigma_{\parallel}$, $\sigma_{\perp}$ and $\sigma_{\perp}^{R}$ represent the conductivity of longitudinal, and two components of transversely polarised spin currents, respectively, and $\hat{m}$ denotes the unit vector along the magnetization of the FM \cite{davidson2020perspectives}. $\vec{Q}_z$ and $\hat{m}$ can be expressed by their components along x, y and z direction. \\
\begin{equation}\label{eq:2} 
 \vec{Q}_z = Q_{zx}\hat{x} + Q_{zy}\hat{y} + Q_{zz}\hat{z}
\end{equation}
\begin{equation}\label{eq:3}
\hat{m} = m_{x}\hat{x} + m_{y}\hat{y} + m_{z}\hat{z}
\end{equation}
In our model, the applied charge current is directed along the x-axis, related by the electric field $\vec{J}_{C}=\frac{1}{\rho_{FM}} E \hat{x}$. The magnetization lies in xy plane in our experiment which makes $m_{z}$=0. The spin Hall like angles are correlated with the spin current conductivity as: $\sigma_{\parallel /\perp /\perp^{R}} = \frac{1}{\rho_{FM}}\theta_{\parallel /\perp /\perp^{R}}$ = $\theta_{\parallel /\perp /\perp^{R}} J_{C}/E$. The expression for spin current is modified as.
\begin{equation}\label{eq:4}
\vec{Q}_z = [\theta_{\parallel}(\hat{m}\cdot\hat{y})\hat{m} + \theta_{\perp}[\hat{m}\times (\hat{y}\times \hat{m})] + \theta_{\perp}^{R} (\hat{m}\times \hat{y})]J_{C}
\end{equation}

\begin{equation}\label{eq:5}
\vec{Q}_z = [\theta_{\parallel}cos(\phi)\hat{m} + \theta_{\perp}sin(\phi)\hat{m}_{\perp} + \theta_{\perp}^{R}sin(\phi)\hat{z}]J_{C}
\end{equation}

\noindent Where, $\hat{m}_\perp$ is unit vectors perpendicular to $\hat{m}$. Our consideration involves the presence of different magnetization in bulk and surface layers, denoted by $\vec{m}$ and $\vec{M}$ respectively. Spin current generated at the surface is symbolized by $\vec{Q}_{z}$ which propagates in the +z direction and reaches the  bulk layer within the FM:\\

\begin{equation}\label{eq:6}
\vec{Q}_z = [\theta_{\parallel}(\hat{M}\cdot\hat{y})\hat{M} + \theta_{\perp}\hat{M}\times (\hat{y}\times\hat{M}) + \theta_{\perp}^{R} (\hat{M}\times\hat{y})]J_{C}\\
\end{equation}

\noindent Now, after few steps of simplification, $\vec{Q}_{z}$ can be simplified in the form.

\begin{equation}\label{eq:7}
\vec{Q}_z = [(\theta_{\parallel}-\theta_{\perp})M_{x}M_{y}\hat{x}  + (\theta_{\parallel} M^{2}_{y} + \theta_{\perp}M_{x}^2)\hat{y} + \theta_{\perp}^{R} M_{x} \hat{z}]J_{C}
\end{equation}

\noindent Similarly, the spin current $\vec{q}_{z}$ generated within the bulk layer with magnetization $\vec{m}$ that flows in the -z direction to the surface can be expressed as:

\begin{equation}\label{eq:8}
\vec{q}_z = [-(\theta_{\parallel}-\theta_{\perp})m_{x}m_{y}]\hat{x}  - [\theta_{\parallel} m^{2}_{y} + \theta_{\perp}m_{x}^2]\hat{y} - \theta_{\perp}^{R} m_{x} \hat{z}]J_{C}  
\end{equation}

\noindent According to Onsagar's reciprocal relation \cite{Onsagar_1931}, a spin current generated at the surface $\vec{Q}_{z}$ along z-axis must also give rise to a charge current within the bulk layer:
\begin{equation}\label{eq:9}
\noindent \vec{J}_e(I \rightarrow B) = \theta_{\parallel}[(\hat{m}.\vec{Q}_{z})\hat{m}\times \hat{z}] + \theta_{\perp}[(\hat{m}\times (\vec{Q}_{z} \times \hat{m}))\times \hat{z}] + \theta_{\perp}^{R} [(\hat{m}\times \vec{Q}_{z})\times\hat{z}]
\end{equation}
In our experiment the charge current is generated in transverse direction due to the spin current. The y-component of $J_e(I \rightarrow B)$ is expressed in a simplified form:

\begin{multline}\label{AISHE:eq:10}
{J}_e^{y}(I \rightarrow B) = [[(\theta_{\parallel}-\theta_{\perp})M_{x}M_{y}][-(\theta_{\parallel}-\theta_{\perp})m_{x}^{2} -\theta_{\perp}] \\+ [\theta_{\parallel} M^{2}_{y} + \theta_{\perp}M_{x}^2][-(\theta_{\parallel}-\theta_{\perp})m_{x}m_{y}] -
(\theta_{\perp}^{R})^{2} M_{x}m_{y}]J_{C} \\
\end{multline}

\noindent Similarly, a spin current $\vec{q}_{z}$ generated within the bulk layer that flows along -z must generate a charge current across surface layer:

\begin{multline}\label{AISHE::eq:11}
{J}_e^{y}(B \rightarrow I) = [[-(\theta_{\parallel}-\theta_{\perp})m_{x}m_{y}][(\theta_{\parallel}-\theta_{\perp})M_{x}^{2} +\theta_{\perp}] \\+ [\theta_{\parallel} m^{2}_{y} + \theta_{\perp}m_{x}^2][-(\theta_{\parallel}-\theta_{\perp})M_{x}M_{y}] -
(\theta_{\perp}^{R})^{2} m_{x}M_{y}]J_{C} \\
\end{multline}
As per our model in the single layer Py, the magnetization $\hat{m}$ and $\hat{M}$ can be expressed as, $\hat{m} = [cos({\phi})\hat{x} + sin({\phi})\hat{y}]$ and  $\hat{M} = [-sin({\alpha})\hat{x} + cos({\alpha})\hat{y}]$. Asymmetric contribution in Eq. \ref{AISHE:eq:10} and \ref{AISHE::eq:11} comes from the last terms - $(\theta_{\perp}^{R})^{2}M_{x}m_{y}J_{C}$ and -$(\theta_{\perp}^{R})^{2}m_{x}M_{y}J_{C}$. The total asymmetric contribution $J_{e(asym)}^{y}$ can be expressed as:\\
\begin{multline}\label{AISHE::eq:12}
J_{e(asym)}^{y} = -(\theta_{\perp}^{R})^{2}M_{x}m_{y}J_{C} -(\theta_{\perp}^{R})^{2}m_{x}M_{y}J_{C}\\
= -(\theta_{\perp}^{R})^{2}J_{C}( m_{x}M_{y} + M_{x}m_{y})\\
= -(\theta_{\perp}^{R})^{2}J_{C}[cos(\phi)cos(\alpha) - sin(\alpha)sin(\phi)]\\
= -(\theta_{\perp}^{R})^{2}J_{C}cos(\phi+\alpha)\\
\end{multline}

\noindent To stop the flow of this charge, an open circuit voltage is developed which is measured and can be described using Eq. \ref{AISHE::eq:12} as:
\begin{equation}\label{AISHE::eq:13}
V_{xy} = wE_{y} = w\rho_{FM} J_{e(asym)}^{y} = -w\rho_{FM} J_{C}(\theta^{R}_{\perp})^{2}cos(\phi+\alpha)    
\end{equation}
Where, $w$ is the width of the Hall bar, $J_{C}$ is the applied current along x. Correspondingly, resistance $R_{xy}$ is expressed as follows.
\begin{equation}\label{AISHE::eq:14}
R_{xy} = V_{xy}/(J_{C}A) = -\frac{(\theta^{R}_{\perp})^{2} \rho_{FM} }{t} cos(\phi+\alpha)
\end{equation}
$R_{xy}$ peaks at a point where $(\alpha+\phi)=180^{0}$, denoted by $\Delta R^{asym}_{xy}(max)$. From equation \ref{AISHE::eq:14}, $\theta^{R}_{\perp}$ is described as.
\begin{equation}\label{AISHE:eq:15}
\theta^{R}_{\perp} = \sqrt{\frac{\Delta R_{xy}^{asym}(max)t}{\rho_{FM}}}
\end{equation}

\begin{table}[h!]
\begin{center}
\begin{tabular}{ | m{3em} | m{2cm}| m{4cm} | m{3cm} | m{2cm} | m{3cm} |} 
  \hline
  Device & $\alpha$ & $\Delta^{sym}_{xy} (\Omega)$ & $\Delta^{asym}_{xy} (m \Omega)$ & $\phi$ & $\theta^{R}_{\perp}$\\ 
  \hline
  D1 & $90^{0}$ & $0.057\pm7.48\times 10^{-5}$ & $5.80\pm0.032$ & $90^{0}$ & $0.016\pm0.001$\\ 
  \hline
  D2 & $60^{0}$ & $0.063\pm5.02\times 10^{-5}$ & $5.92\pm0.007$ & $120^{0}$ & $0.016\pm5.46\times10^{-4}$ \\ 
  \hline
  D3 & $45^{0}$ & $0.064\pm7.04\times 10^{-5}$ & $5.96\pm0.008$ & $135^{0}$ & $0.016\pm5.71\times10^{-4}$\\
  \hline
  D4 & $30^{0}$ & $0.066\pm6.11\times 10^{-5}$ & $6.12\pm0.005$ & $150^{0}$ & $0.016\pm4.61\times10^{-4}$\\
  \hline
  D5 & $0^{0}$ & $0.060\pm6.39\times 10^{-5}$ & $6.55\pm0.039$ & $180^{0}$ & $0.016\pm0.001$\\
  \hline  
\end{tabular}
\end{center}
\caption{The results from 1w measurement in obliquely deposited Py devices are summarised. The parameters such as $\alpha$, $\Delta R ^{sym}_{xy}$, $\Delta R^{asym}_{xy}$, $\phi$, and $\theta^{R}_{\perp}$ are listed for devices D1-D5.}
\end{table}

\clearpage
\noindent \textbf{Note 4. 2w field scans for devices D1 and D5}:\\
\begin{figure}[h!]
\centering
\includegraphics[width=1\textwidth]{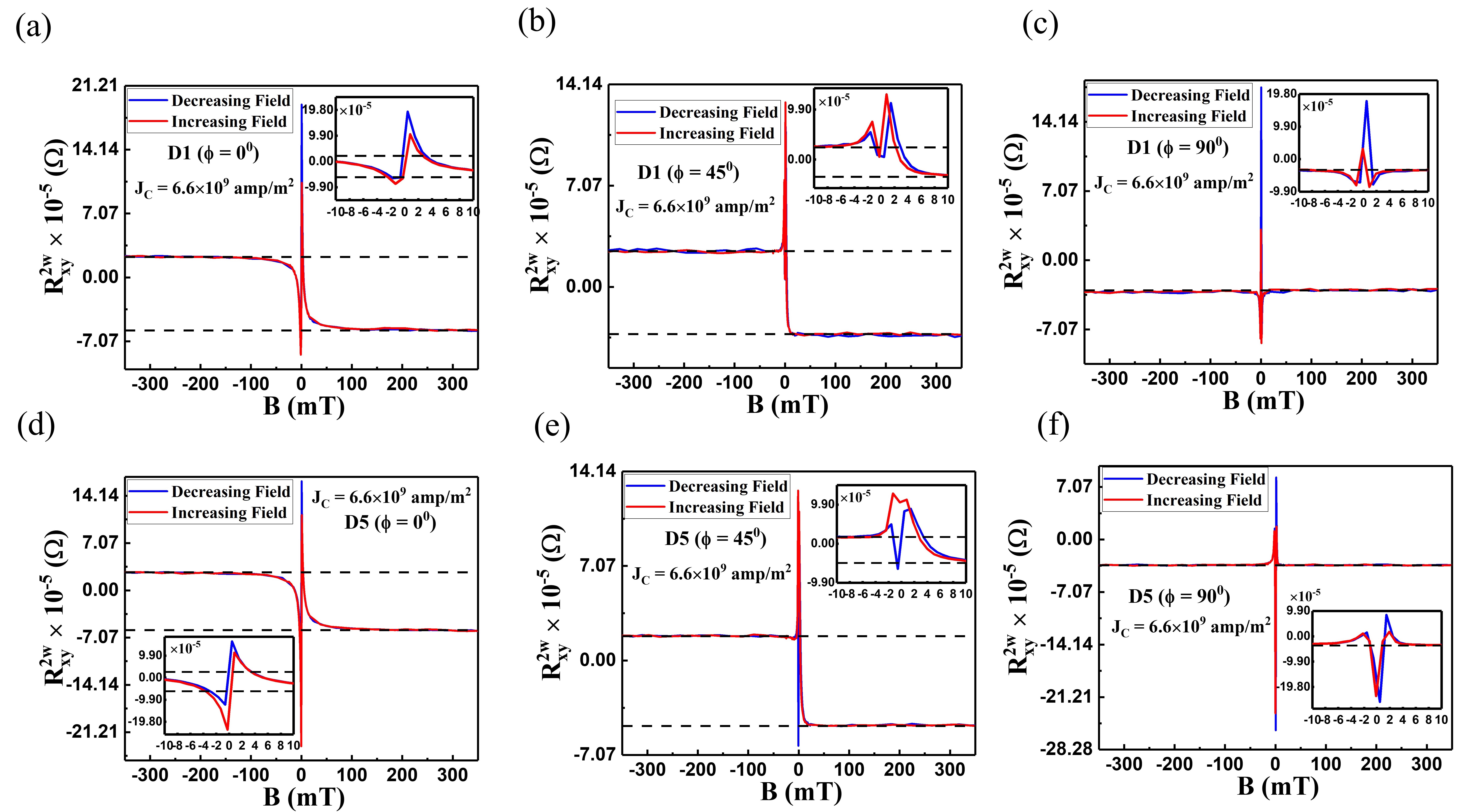}
\caption{(a)-(f) Shows the magnetic field (B) dependence of $R^{2w}_{xy}$ for devices D1 and D5 at angles $\phi$ = $0^{0}$, $45^{0}$, and $90^{0}$. B is swept from + 350 mT to -350 mT and vice versa, with blue, red lines indicate decreasing, increasing fields, respectively (insets: shows the graphs near zero field ($\pm$ 10 mT) with same y-axis). The dotted lines serve as guideline to indicate the saturation region in those graphs.}
\label{SM_Figure 5}
\end{figure}

\noindent \textbf{Note 5. Calculation for 2w transport measurement}:\\

\noindent Similar treatment as 1w formulation can be used to present the toy model for 2w. When a ac current is applied along the permalloy Hall bar, due to self-induced SOT the spin current exerts torques within the FM itself. The Spin current generated in the bulk layer creates torques at the surface with magnetization $\hat{M}$ are $(\hat{M}\times\vec{q}_{z})$ and $[\hat{M}\times(\hat{M}\times\vec{q}_{z})]$ \cite{avci2014interplay,Ohno_2014}. These two torques can be generalized as:
\begin{flalign*}
\noindent (\hat{M}\times\vec{q}_{z})= [(M_{x}\hat{x}+M_{y}\hat{y}) \times (-(\theta_{\parallel}-\theta_{\perp})m_{x}m_{y}\hat{x}  - (\theta_{\parallel} m^{2}_{y} + \theta_{\perp}m_{x}^2)\hat{y} - \theta_{\perp}^{R} m_{x} \hat{z})]J_{C}\\
\noindent = \theta_{\perp}^{R}m_{x}J_{C}\hat{M}_{\perp} + [M_{y}(\theta_{\parallel}-\theta_{\perp})m_{x}m_{y}-(\theta_{\parallel}m^{2}_{y}+\theta_{\perp}m^{2}_{x})M_{x}]\hat{z}J_{C}\\
\end{flalign*}

\begin{equation}\label{AISHE:eq:16}
(\hat{M}\times\vec{q}_{z}) =\hat{M}\times[-\theta_{\perp}^{R}m_{x}J_{C}\hat{z}+\hat{M}_{\perp}(M_{y}(\theta_{\parallel}-\theta_{\perp})m_{x}m_{y}-(\theta_{\parallel}m^{2}_{y}+\theta_{\perp}m^{2}_{x})M_{x})J_{C}]\\
\end{equation}
and
\begin{equation}\label{AISHE:eq:17}
\begin{split}
[\hat{M}\times(\hat{M}\times\vec{q}_{z})]= \theta_{\perp}^{R}m_{x}J_{C}(\hat{M}\times\hat{M}_{\perp})+(\hat{M}\times\hat{z})[(M_{y}(\theta_{\parallel}-\theta_{\perp})m_{x}m_{y}-(\theta_{\parallel}m^{2}_{y}+\theta_{\perp}m^{2}_{x})M_{x})J_{C}]\\
=\hat{M}\times[\theta_{\perp}^{R}m_{x}J_{C}\hat{M}_{\perp}+(M_{y}(\theta_{\parallel}-\theta_{\perp})m_{x}m_{y}-(\theta_{\parallel}m^{2}_{y}+\theta_{\perp}m^{2}_{x})M_{x})J_{C}\hat{z}]
\end{split}
\end{equation}

\noindent where, $\hat{M}\times\hat{z}=-\hat{M}_{\perp}$ and $\hat{M}\times\hat{M}_{\perp}=\hat{z}$. The torques $(\hat{M}\times\vec{q}_{z})$ and $[\hat{M}\times(\hat{M}\times\vec{q}_{z})]$ are equivalent to the current induced fields $\vec{q}_{z}$ and $(\hat{M}\times\vec{q}_{z})$. Now considering $\hat{z}$ and $\hat{M}_{\perp}$ to be $\theta$ and $\phi$ respectively, $B_{\theta}^{'}$ and $B_{\phi}^{'}$ can be described as:

\begin{equation}\label{AISHE:eq:18}
B_{\theta}^{'} = -\frac{\hbar}{2e}\frac{\theta^{R}_{\perp}J_{C}m_{x}}{M_{S}\lambda}
\end{equation}

\begin{equation}\label{AISHE:eq:19}
B_{\phi}^{'} = \frac{\hbar}{2e}\frac{\theta^{R}_{\perp}J_{C}m_{x}}{M_{S}\lambda}
\end{equation}
\noindent Where, $B^{'}_{\theta}$ and $B^{'}_{\phi}$ represent the damp-like and field-like fields due to spin current in bulk. Similarly, the spin current generated at the surface layer creates torques at the bulk with magnetization $\hat{m}$ are $(\hat{m}\times\vec{Q}_{z})$ and $[\hat{m}\times(\hat{m}\times\vec{Q}_{z})]$. These two torques can be generalized as:

\begin{flalign}\label{AISHE:eq:20}
(\hat{m}\times\vec{Q}_{z})= [(m_{x}\hat{x}+m_{y}\hat{y}) \times ((\theta_{\parallel}-\theta_{\perp})M_{x}M_{y}\hat{x}  + (\theta_{\parallel} M^{2}_{y} + \theta_{\perp}M_{x}^2)\hat{y} +\theta_{\perp}^{R} M_{x} \hat{z})]J_{C} \nonumber \\
=\hat{m}\times[\theta_{\perp}^{R}M_{x}J_{C}\hat{z}+\hat{m}_{\perp}((\theta_{\parallel} M^{2}_{y} + \theta_{\perp}M_{x}^2)m_{x}-m_{y}(\theta_{\parallel}-\theta_{\perp})M_{x}M_{y})J_{C}]\\ \nonumber
\end{flalign}
and
\begin{flalign}\label{AISHE:eq:21}
\hat{m}\times(\hat{m}\times\vec{Q}_{z})= \hat{m}\times[\theta_{\perp}^{R}M_{x}J_{C}(\hat{m}\times\hat{z})+(\hat{m}\times\hat{m}_{\perp})[(\theta_{\parallel}M^{2}_{y}+\theta_{\perp}M^{2}_{x})m_{x}-m_{y}(\theta_{\parallel}-\theta_{\perp})M_{x}M_{y})J_{C}] \nonumber\\
=\hat{m}\times[-\theta_{\perp}^{R}M_{x}J_{C}\hat{m}_{\perp}+\hat{z}[(\theta_{\parallel}M^{2}_{y}+\theta_{\perp}M^{2}_{x})m_{x}-m_{y}(\theta_{\parallel}-\theta_{\perp})M_{x}M_{y})J_{C}]\\ \nonumber
\end{flalign}

\noindent where, $\hat{m}\times\hat{z}=-\hat{m}_{\perp}$ and $\hat{m}\times\hat{m}_{\perp}=\hat{z}$. The torques $(\hat{m}\times\vec{Q}_{z})$ and $[\hat{m}\times(\hat{m}\times\vec{Q}_{z})]$ are equivalent to the fields $\vec{Q}_{z}$ and $(\hat{m}\times\vec{Q}_{z})$. $B_{\theta}^{''}$ and $B_{\phi}^{''}$ can be described as: 

\begin{equation}\label{AISHE:eq:22}
B_{\theta}^{''} = \frac{\hbar}{2e}\frac{J_{C}}{M_{S}\lambda}[(\theta_{\parallel}M^{2}_{y}+\theta_{\perp}M^{2}_{x})m_{x}-m_{y}(\theta_{\parallel}-\theta_{\perp})M_{x}M_{y})]
\end{equation}

\begin{equation}\label{AISHE:eq:23}
B_{\phi}^{''} = \frac{\hbar}{2e}\frac{J_{C}}{M_{S}\lambda}[(\theta_{\parallel}M^{2}_{y}+\theta_{\perp}M^{2}_{x})m_{x}-m_{y}(\theta_{\parallel}-\theta_{\perp})M_{x}M_{y})]
\end{equation}

\noindent Where, $B^{''}_{\theta}$ and $B^{''}_{\phi}$ represent the damp-like and field-like fields due to spin current in surface. The net current-induced field (asymmetric) can be represented by:

\begin{equation}\label{AISHE:eq:24}
B_{\theta} = \frac{\hbar}{2e}\frac{J_{C}}{M_{S}\lambda}[m_{x}(\theta_{\parallel}M^{2}_{y}+\theta_{\perp}M^{2}_{x}-\theta^{R}_{\perp})-m_{y}(\theta_{\parallel}-\theta_{\perp})M_{x}M_{y})]
\end{equation}

\begin{equation}\label{AISHE:eq:25}
B_{\phi} = \frac{\hbar}{2e}\frac{J_{C}}{M_{S}\lambda}[m_{x}(\theta_{\parallel}M^{2}_{y}+\theta_{\perp}M^{2}_{x}+\theta^{R}_{\perp})-m_{y}(\theta_{\parallel}-\theta_{\perp})M_{x}M_{y})]
\end{equation}

\noindent The net symmetric contribution from the current induced fields can be represented by:

\begin{equation}\label{AISHE:eq:26}
B_{\theta}(sym) = \frac{\hbar}{2e}\frac{J_{C}}{M_{S}\lambda}[-M_{x}(\theta_{\parallel}m^{2}_{y}+\theta_{\perp}m^{2}_{x}-\theta^{R}_{\perp})+M_{y}(\theta_{\parallel}-\theta_{\perp})m_{x}m_{y})]
\end{equation}

\begin{equation}\label{AISHE:eq:27}
B_{\phi}(sym) = \frac{\hbar}{2e}\frac{J_{C}}{M_{S}\lambda}[-M_{x}(\theta_{\parallel}m^{2}_{y}+\theta_{\perp}m^{2}_{x}+\theta^{R}_{\perp})+M_{y}(\theta_{\parallel}-\theta_{\perp})m_{x}m_{y})]
\end{equation}\\

\noindent \textbf{Calculation of resistance for 2w}:\\

\noindent When an ac current with a density of $J_{C}=J_{0}sin(wt)$ is applied in a FM, the transverse voltage reads: $V_{xy}(t)$ = $R_{xy}(t)J_{0}sin(wt)A$. Correspondingly, resistance is given by: $R_{xy}(t) = R_{xy}(B+B_{IF})$, where, $B$ denotes the external magnetic field, $B_{IF}$ represents the current induced fields including damp-like, field-like and Oersted fields. These fields tend to deviate the magnetization of FM. When the oscillation is small, resistance can be expanded into the first order as. 
\begin{equation}\label{AISHE:2w:eq:R1}
R_{xy}(t) = R(B) + \frac{dR_{xy}}{dB_{IF}}B_{IF}sin(wt)
\end{equation}
By inserting this equation in the transverse voltage expression, $V_{xy}(t)$ can be expanded.
\begin{equation}\label{AISHE:2w:eq:R2}
V_{xy}(t) = J_{0}A[R^{0}_{xy} + R^{1w}_{xy}sin(wt) + R^{2w}_{xy}cos(2wt)]
\end{equation}
Where, $R^{0}_{xy}$ = $\frac{1}{2} \frac{dR_{xy}}{dB_{IF}}$, $R^{1w}_{xy}$ = $R_{xy}(B)$, $R^{2w}_{xy}$ = -$\frac{1}{2} \frac{dR_{xy}}{dB_{IF}}$ represent the zero, 1st, and 2nd order components of harmonics. 1st order component is equivalent to dc measurements and is expressed as.
\begin{equation}\label{AISHE:2w:eq:R3}
R^{1w}_{xy} = R_{AHE} cos(\theta)+R_{PHE} sin^{2}(\theta) sin(2\phi)
\end{equation}
Similarly, 2nd order component is given by.
\begin{equation}\label{AISHE:2w:eq:R4}
R^{2w}_{xy} = \frac{dR^{1w}_{xy}}{d\theta_{B}}\frac{B_{\theta}}{B} + \frac{dR^{1w}_{xy}}{d\phi_{B}}\frac{B_{\phi}}{B} + J_{0}A\alpha\nabla Tcos(\phi)
\end{equation}

\noindent $\frac{dR^{1w}_{xy}}{d\theta_{B}} = -R_{AHE} sin(\theta) = -R_{AHE} (\theta = \pi/2)$\\
$\frac{dR^{1w}_{xy}}{d\phi_{B}} = 2R_{PHE} cos(2\phi)sin^{2}(\theta)=2R_{PHE} cos(2\phi)  (\theta = \pi/2)$\\
$R^{2w}_{xy}$ can be modified into the expression:
\begin{equation}\label{AISHE:2w:eq:R5}
R^{2w}_{xy} = -R_{AHE} \frac{B_{\theta}}{B+B_{de}} + 2R_{PHE}cos(2\phi)\frac{B_{\phi}}{B} + J_{0}A\alpha\nabla Tcos(\phi)
\end{equation}

\begin{equation}\label{AISHE:2w:eq:R6}
R^{2w}_{xy} = -\frac{R_{AHE}}{B+B_{de}} (a^{'}cos(\phi)-c^{'}sin(\phi))+ \frac{2R_{PHE}cos(2\phi)}{B}(b^{'}cos(\phi)-d^{'}sin(\phi)) \\+ J_{0}A\alpha\nabla Tcos(\phi)
\end{equation}

\begin{equation}\label{AISHE:2w:eq:all_coef}
R^{2w}_{xy}=a cos(\phi) +b cos(2\phi)cos(\phi)+c sin(\phi) + d cos(2\phi)sin(\phi) + e + f sin(2\phi) 
\end{equation}
\clearpage
\noindent \textbf{Damp-like and field-like coefficients for device D5:}
\begin{figure}[h!]
\centering
\includegraphics[width=1\textwidth]{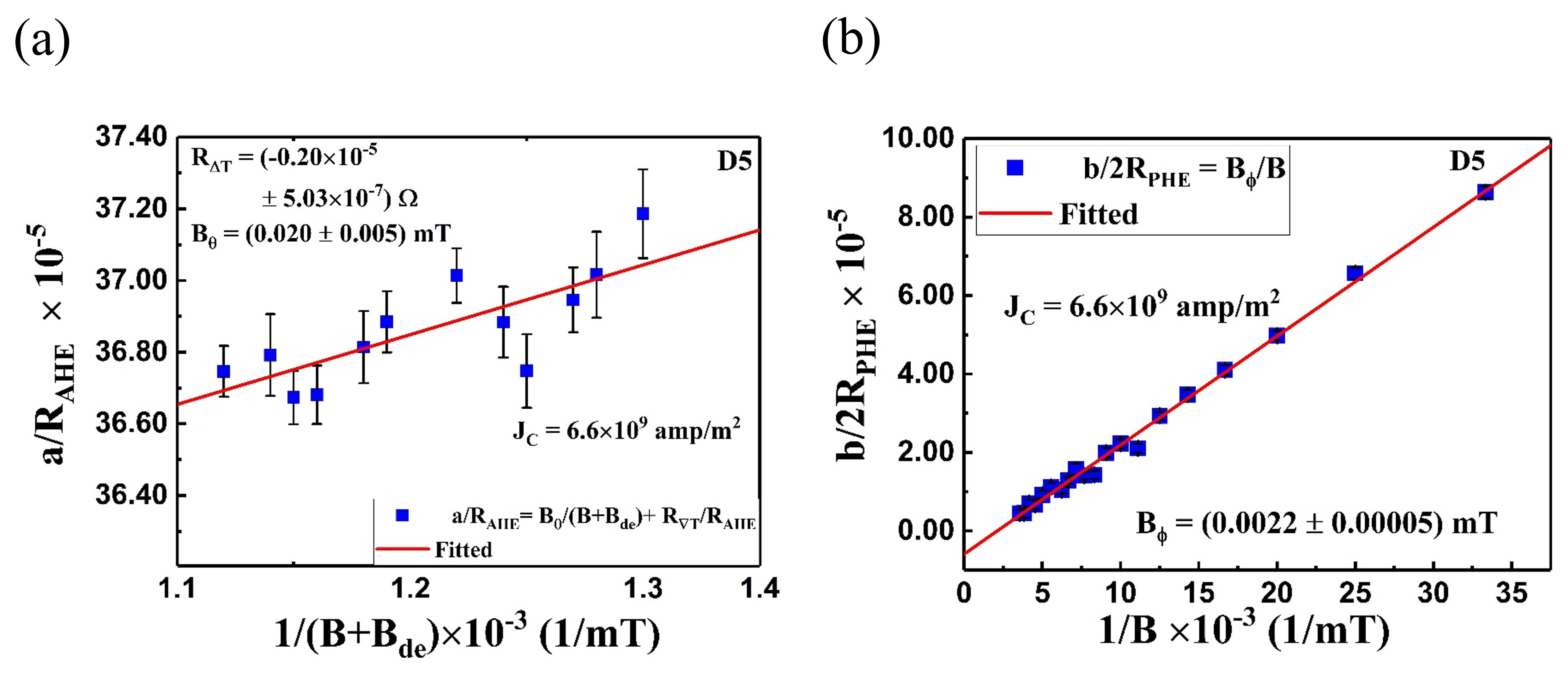}
\caption{(a) Shows the behaviour of $a/R_{AHE}$ vs $\frac{1}{(B+B_{de})}$ for device D5. The curve is fitted with a straight line, the constant term yields the ANE term $R_{\nabla T}$, while the slope evaluates $B_{\theta}$, (b)) Illustrates $b/2R_{PHE}$ as a function of $\frac{1}{B}$ for device D5. From the straight line fit, $B_{\phi}$ can be evaluated. The blue squares represent the measured data points, and the red solid lines depict the fitted straight lines.}
\label{SM_Figure 6}
\end{figure}

\begin{table}[h!]
\begin{center}
\begin{tabular}{ | m{3em} | m{2.5cm}| m{2.0cm} | m{3.0cm} | m{2.5cm} | m{2cm} |}  
  \hline
  Device & $B_{\theta}(mT)$ & $B_{\phi}(mT)$ & $R_{\nabla T}(\Omega)$ & $\xi_{\theta}$ & $\xi_{\phi}$\\ 
  \hline
  D1 & $0.026 \pm 0.005$ & $0.0014\pm 0.00004$ & $-3.12\times 10^{-5} \pm 4.54\times 10^{-7}$ & $0.024 \pm 0.003$ & $0.0014\pm 0.00004$\\ 
  \hline
  D2 & $0.024 \pm 0.001$ & $0.00085\pm 0.00004$ & $-2.73\times 10^{-5} \pm 1.36\times 10^{-7}$ & $0.022 \pm 0.0009$ & $0.0008 \pm 0.00004$ \\ 
  \hline
  D3 & $0.010 \pm 0.003$ & $0.0012\pm 0.00003$ & $-2.44\times 10^{-5} \pm 2.44\times 10^{-7}$ & $0.009 \pm 0.003$ & $0.0011 \pm 0.00003$\\
  \hline
  D4 & $0.019 \pm 0.002$ & $0.0015\pm 0.00005$ & $-0.87\times 10^{-5} \pm 1.20\times 10^{-7}$ & $0.018 \pm 0.002$ & $0.0014 \pm 0.00005$\\
  \hline
  D5 & $0.020 \pm 0.005$ & $0.0024\pm 0.00005$ & $-0.2\times 10^{-5} \pm 5.03\times 10^{-7}$ & $0.019 \pm 0.005$ & $0.0022 \pm 0.00005$\\
  \hline  
\end{tabular}
\end{center}
\caption{The summary of the 2w harmonic measurements are listed for devices D1-D5. The evaluated parameters including $B_{\theta}$, $B_{\phi}$, $R_{\nabla T}$, $\xi_{\theta}$, and $\xi_{\phi}$ are organized.}
\label{table:2w}
\end{table}

\clearpage

\bibliography{references}

\end{document}